\makeatletter\def\input@path{
{Figures/}
}\makeatother%

\documentclass[aps,pra,onecolumn,nofootinbib,superscriptaddress,
amsmath,amssymb,showkeys]{revtex4-2}

\usepackage{chemformula,chemfig}
\setchemfig{atom style={scale=0.4}, double bond sep = 4pt}
\usepackage[version=4,arrows=pgf-filled,textfontname=sffamily,
mathfontname=mathsf]{mhchem}
\usepackage{microtype}  
\usepackage[detect-all=true,group-minimum-digits=4]{siunitx}
\usepackage{tabularx, longtable}%
\usepackage{booktabs, multirow,xcolor,cleveref,url}

\newcommand{\deltaest}{$\Delta E_{\text{ST}}$}

\DeclareSIUnit\calorie{cal}
\DeclareSIUnit\kcal{\kilo\calorie}
\DeclareSIUnit\kcal{Kcal}
\DeclareSIUnit\atomicunit{a.u.}

\begin{document}
\author{Jean-Pierre Tchapet Njafa}
\email{jean-pierre.tchapet@facsciences-uy1.cm}
\affiliation{Department of Physics, Faculty of Science, University of Yaounde I, P.O.Box
812, Yaounde, Cameroon}

\author{Elvira Vanelle Kameni Tcheuffa}
\affiliation{Department of Physics, Faculty of Science, University of Yaounde I, P.O.Box
812, Yaounde, Cameroon}

\author{Aissatou Maghame}
\affiliation{Department of Physics, Faculty of Science, University of Yaounde I, P.O.Box
812, Yaounde, Cameroon}

\author{S. G. Nana Engo}
\email{serge.nana-engo@facsciences-uy1.cm}
\affiliation{Department of Physics, Faculty of Science, University of Yaounde I, P.O.Box
812, Yaounde, Cameroon}

\title{Validation of Semi-Empirical xTB Methods for High-Throughput Screening of TADF Emitters: A \num{747}-Molecule Benchmark Study}



\begin{abstract}
Thermally activated delayed fluorescence (TADF) emitters are essential for next-generation, high-efficiency organic light-emitting diodes (OLEDs), yet their rational design is hampered by the high computational cost of accurate excited-state predictions. Here, we present a comprehensive benchmark study validating semi-empirical extended tight-binding (xTB) methods—specifically sTDA-xTB and sTD-DFT-xTB—for the high-throughput screening of TADF materials. Using an unprecedentedly large dataset of \num{747} experimentally characterized emitters, our framework demonstrates a computational cost reduction of over \qty{99}{\percent} compared to conventional TD-DFT, while maintaining strong internal consistency between methods (Pearson $r \approx \num{0.82}$ for \deltaest), validating their utility for relative molecular ranking. Validation against \num{312} experimental \deltaest values reveals a mean absolute error of approximately \qty{0.17}{\electronvolt}, a discrepancy attributed to the vertical approximation inherent to the HTS protocol, underscoring the methods' role in screening rather than quantitative prediction. Through large-scale data analysis, we statistically validate key design principles, confirming the superior performance of Donor-Acceptor-Donor (D-A-D) architectures and identifying an optimal D-A torsional angle range of \qtyrange{50}{90}{\degree} for efficient TADF. Principal Component Analysis reveals that the complex property space is fundamentally low-dimensional, with three components capturing nearly \qty{90}{\percent} of the variance. This work establishes these semi-empirical methods as powerful, cost-effective tools for accelerating TADF discovery and provides a robust set of data-driven design rules and methodological guidelines for the computational materials science community.
\end{abstract}

\maketitle

\section{Introduction}
\label{sec:introduction}

Organic Light-Emitting Diodes (OLEDs) utilizing Thermally Activated Delayed Fluorescence
(TADF) emitters have emerged as a third generation of optoelectronic devices, promising
near 100\% internal quantum efficiency by effectively harvesting both singlet and triplet
excitons via Reverse Intersystem Crossing (RISC) \cite{Uoyama2012, Adachi2014TADF,
Liu2018}. The key photophysical prerequisite for efficient TADF is a small singlet-triplet
energy gap ($\Delta E_{\text{ST}} \le \qty{0.3}{\electronvolt}$), typically achieved
through molecular designs that promote a spatial separation of the highest occupied
molecular orbital (HOMO) and the lowest unoccupied molecular orbital (LUMO), leading to
excited states with substantial Charge-Transfer (CT) character \cite{Dias2017, Chen2018}.

The rational computational design and discovery of novel TADF emitters, however, face significant theoretical challenges \cite{Olivier2018, Gomez-Bombarelli2016}. Conventional \textit{ab initio} methods, such as high-level state function theory (e.g., SCS-CC2 or STEOM-DLPNO-CCSD), often required for quantitative accuracy, especially for complex Multi-Resonance (MR-TADF) systems) \cite{Kang2024, Wu2025}, are computationally prohibitive for the high-throughput screening (HTS) of the vast chemical space relevant to OLED materials \cite{Olivier2017a}. Time-Dependent Density Functional Theory (TD-DFT), while more efficient, suffers from well-documented limitations, particularly its tendency to underestimate the energy of highly delocalized CT states and its known failures in contexts requiring multireference character or the consideration of environmental polarization effects \cite{Dreuw2004, Mewes2018, Jacquemin2009, Froitzheim2022}. Furthermore, calculating the precise \deltaest in solution, a critical metric for TADF performance, ideally requires computationally expensive approaches like the State-Specific Polarizable Continuum Model (SS-PCM) or Restricted Open-shell Kohn-Sham (ROKS) methods, which account for the differential stabilization of $S_1$ and $T_1$ states in polar environments \cite{Loos2020STGABS27, Laurent2013TDDFT}.

To address this fundamental dilemma—the need for robust predictions across thousands of molecules constrained by the cost and accuracy shortcomings of current methods—Simplified Quantum Chemistry (sQC) approaches have gained traction \cite{bannwarth2021extended}. In this context, the combination of eXtended Tight-Binding (xTB) methods with simplified response theory offers a compelling pathway forward \cite{Grimme2016, Grimme2021, Wergifosse2024}. Specifically, the simplified Tamm-Dancoff approximation (sTDA) and simplified TD-DFT (sTD-DFT) methods, when coupled with the efficient xTB Hamiltonian (sTDA-xTB and sTD-DFT-xTB), provide a highly scalable framework capable of treating systems with hundreds to thousands of atoms \cite{Grimme2016, Grimme2013}. This capability is achieved by approximating two-electron integrals and truncating the configuration interaction (CI) space, leading to computational cost reductions often exceeding 99\% compared to conventional TD-DFT calculations.

This work presents a rigorous, large-scale validation of the semi-empirical sTDA-xTB and
sTD-DFT-xTB methods as tools for the high-throughput virtual screening (HTVS) of TADF
emitters. We leverage a uniquely extensive dataset comprising \num{747} experimentally
characterized TADF molecules \cite{Huang_2024}, moving significantly beyond the
limitations of small datasets previously assessed
\cite{Moral2015, deSilva2019TADF}. In doing so, we compare the performance of
these methods against each other and experimental data, assess their ability to capture
solvent effects relevant to device environments, and evaluate the computational
cost-benefit ratio relative to conventional approaches.

We systematically evaluate the fidelity of sTDA-xTB and sTD-DFT-xTB by comparison against
experimental data, including \num{312} experimental \deltaest values and \num{213}
emission wavelengths ($\lambda_{\text{PL}}$) \cite{Huang_2024}. This massive validation
effort establishes the strengths and limitations of sQC methods for predicting key TADF
properties. We confirm the strong internal consistency of the sTDA-xTB and sTD-DFT-xTB
variants ($\text{Pearson } r \approx 0.82$ for \deltaest), demonstrating their equivalence
for the critical task of relative molecular ranking essential for HTS, despite
acknowledging known quantitative inaccuracies for absolute values ($\text{MAE } \sim
\qty{0.17}{\electronvolt}$ against experiment for \deltaest) \cite{Zhao2021}. We
explicitly investigate the impact of the approximations inherent to the sQC approach,
including the hybrid use of GFN2-xTB geometries (optimized for ground-state properties)
and the implicit solvent model ALPB (a linearized Poisson-Boltzmann model) for calculating
excited-state properties in toluene \cite{Bannwarth2019, Ehlert2021}. We quantify the
statistically significant, albeit modest, influence of solvent effects on photophysical
properties. Our analysis confirms that the computational efficiency achieved enables the
processing of hundreds of molecules rapidly, providing essential benchmarking data and
methodological guidelines for accelerating TADF emitter discovery pipelines
\cite{Perez‐Jimenez2025}. We show that the vast majority of property variance can be
captured by a few components, suggesting a low-dimensional and targetable design space.

The manuscript is organized as follows: Section \ref{sec:Comp_Meth} details the
computational methodology, including the protocols for geometry optimization (GFN2-xTB)
\cite{Grimme2017}, excited-state calculation (sTDA/sTD-DFT-xTB), and the definition of
molecular and photophysical metrics. Section \ref{sec:Res_Disc} presents the results of
the comprehensive benchmarking study, discussing the method consistency, predictive
accuracy against experimental data, and the influence of solvent effects. Finally,
Section \ref{sec:Concl} summarizes the findings and proposes future research directions,
particularly the integration of these validated sQC methods with Machine Learning
techniques for predictive modeling of TADF dynamics.

\section{Computational methods}\label{sec:Comp_Meth}

\begin{figure}[!htbp]
\centering
 \includegraphics{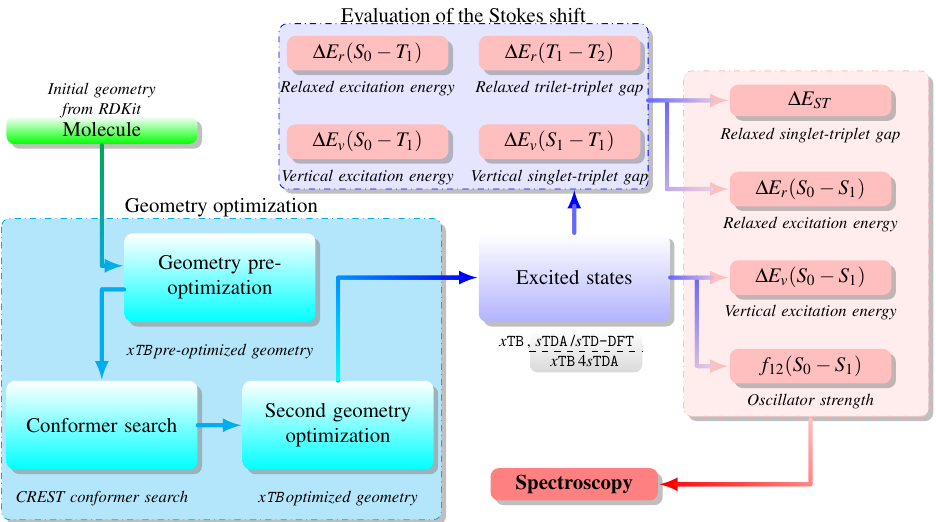}
\caption{Overview of the simulation workflow. Starting with a SMILES string, the code performs conformer search and geometry optimisation via xTB for the singlet ground state $S_0$ and the triplet state $T_1$. It allows the extraction of the relaxed triplet excitation energy. Simplified time-dependent DFT calculation with sTDA/sTD-DFT extracts the vertical singlet-triplet gap, the relaxed triplet-triplet gap, the oscillator strength, the vertical excitation energy and the fluorescence absorption and emission spectra, while incorporating solvent effects to enhance the realism of our simulations. The Stokes shift is evaluated and then allows the relaxed singlet-triplet gap to be estimated.}
\label{fig:workflow}
\end{figure}

The aim of this work is to establish a robust and computationally efficient methodology for the high-throughput virtual screening of Thermally Activated Delayed Fluorescence (TADF) emitters. This protocol integrates accelerated semi-empirical methods from the extended tight-binding (xTB) family for geometry optimization and subsequent excited-state property calculations, enabling the analysis of a large and diverse chemical space.   A schematic representation of the workflow is presented in \Cref{fig:workflow}.

\subsection{Dataset and conformational sampling}\label{sec:geom_opt}

Our benchmark dataset comprises \num{747} TADF emitters extracted from the literature using automated text mining \cite{Huang_2024}. The set spans diverse molecular architectures, including donor-acceptor (D-A), multiple donor-acceptor, and multi-resonance (MR) systems. Initial 3D structures were generated from SMILES strings using RDKit \cite{rdkit}. The complete molecular list is provided in Supporting Information Table S1.

A systematic conformational search for each molecule was performed using the Conformer-Rotamer Ensemble Sampling Tool (CREST version 3.0) \cite{pracht2020automated, crest2} coupled with the GFN2-xTB semi-empirical Hamiltonian using the xTB program (version 6.7.0) \cite{Bannwarth2019, grimme2019exploration}. GFN2-xTB is a second-generation tight-binding quantum chemical method specifically parameterized for accurate molecular structures, conformational energies, and noncovalent interactions, making it highly suitable for describing the complex geometries of TADF emitters \cite{bannwarth2021extended}. The lowest-energy conformer identified by CREST was then subjected to a final, tight geometry optimization at the GFN2-xTB level to obtain the equilibrium structure on the singlet electronic ground state ($S_0$).

\subsection{Excited-state calculations and hybrid protocol}\label{sec:excited_state}

Excited-state properties were calculated using the geometries optimized at the $S_0$ GFN2-xTB level through two highly efficient quantum chemistry methods based on the extended tight-binding (xTB) framework: the simplified Tamm-Dancoff approximation (sTDA-xTB) \cite{Grimme2016} and simplified time-dependent density functional theory (sTD-DFT-xTB) \cite{Grimme2021, Wergifosse2024}. Notably, these excited-state calculations employ a non-self-consistent extended tight-binding Hamiltonian, distinct from the GFN2-xTB Hamiltonian used for ground-state geometry optimization \cite{Grimme2017}. The sTDA-xTB and sTD-DFT-xTB methods utilize a specialized Hamiltonian and an extended atomic orbital (AO) basis set parameterized for robust and rapid prediction of absorption and emission spectra \cite{bannwarth2021extended}.

This hybrid protocol, which adopts GFN2-xTB optimized geometries for a single-point
sTDA/sTD-DFT-xTB calculation, represents a pragmatic approximation enabling the high
computational efficiency necessary for high-throughput screening (HTS). While
excited-state optimized geometries would be preferable for accurately modeling adiabatic
properties \cite{Froitzheim2022, Mewes2018}, this approach preserves the relative rankings
and qualitative trends critical for virtual screening, achieving computational speed-ups
exceeding \qty{99}{\percent} compared to conventional TD-DFT \cite{Konidena_2021}.

We explicitly acknowledge that the GFN2-xTB and the xTB Hamiltonian used for sTDA/sTD-DFT
(denoted as \texttt{xTB4sTDA}) stem from distinct parametrizations and basis sets
\cite{Ma_2019, Wei_2022, Zeng_2019}. The workflow’s key assumption is that the GFN2-xTB
optimized ground-state geometry provides a sufficiently accurate vertical approximation
($S_0$ geometry) to support rapid and reliable single-point excited-state computations.
This separation is justified as the main goal is qualitative trend screening and relative
molecular ranking across hundreds of compounds, delivering a remarkable computational cost
reduction of roughly \qty{99}{\percent} relative to conventional TD-DFT
\cite{Konidena_2021, Gomez-Bombarelli2016}. For precise absolute energies or adiabatic
excited-state geometries, higher-level, more computationally demanding methods would be
necessary \cite{Wang_2021}.

\subsection{Solvent effects modeling}\label{sec:solvation}

To model the influence of the molecular environment, all calculations were performed in
both the gas phase and in solution. We employed the analytical linearized
Poisson-Boltzmann (ALPB) implicit solvation model \cite{Ehlert2021}, with toluene
($\varepsilon = \num{2.38}$) selected as a representative low-polarity solvent commonly
used in experimental studies of TADF emitters. We acknowledge that this model relies on
the ground-state electronic density and does not fully capture dynamic solvent
reorganization effects pertinent to charge-transfer states. For quantitative accuracy,
state-specific (SS-PCM) or restricted open-shell Kohn-Sham (ROKS) methods would be
necessary \cite{Hait2016, Northey2017}. However, the ALPB model provides a computationally
tractable and consistent approach for assessing solvent-induced trends across our large
dataset.

\subsection{Electronic structure and property analysis}\label{sec:analysis}

From the excited-state calculations, we directly obtained the vertical excitation energies
of the lowest singlet ($E_{S_1}$) and triplet ($E_{T_1}$) states, from which the
singlet-triplet energy gap was calculated:
\begin{equation}
    \Delta E_{\text{ST}} = E_{S_1} - E_{T_1}.
\end{equation}
The dimensionless oscillator strength ($f_{S_1}$) for the $S_0 \to S_1$ transition was
also computed as a proxy for the radiative decay rate.

Given that direct geometry optimization of the $S_1$ state is not implemented in the
employed semi-empirical framework, an approximation was necessary to estimate emission
properties. We first calculated the relaxed triplet excitation energy by optimizing the
$T_1$ state geometry at the GFN2-xTB level. The geometric relaxation energy for the
triplet state, $\Delta E_{\text{relax}}(T_1)$, was then assumed to be equal to the Stokes
shift of the $S_1$ state. This physically-grounded approximation allowed for the
estimation of the relaxed $S_1$ energy, and consequently, the emission wavelength
($\lambda_{\text{PL}}$). The detailed equations for this estimation are provided in the
Supporting Information.

To gain physical insight into the electronic structure, particularly the degree of
charge-transfer (CT) character, we analyzed the frontier molecular orbitals. Using the
Multiwfn package \cite{lu2024multiwfn}, we quantified the spatial separation between the
highest occupied (HOMO) and lowest unoccupied (LUMO) molecular orbitals via two key
descriptors: the overlap integral ($S'_{\text{HL}}$) and the centroid distance between
their respective electron densities ($D_{\text{HL}}$) \cite{Etienne2014, LeBahers2011}.
These metrics provide a quantitative measure of the electronic decoupling central to the
TADF mechanism.

\subsection{Validation, statistical analysis, and computational
cost}\label{sec:validation_and_cost}

To validate our computational protocol, we benchmarked the performance of the sTDA-xTB and
sTD-DFT-xTB methods against a large set of experimental data extracted from the
literature. The validation set consisted of \num{312} experimental \deltaest values and
\num{213} experimental emission wavelengths ($\lambda_{\text{PL}}$), standardized to eV
and nm units, respectively. The performance of the semi-empirical methods was rigorously
assessed by comparing their predictions to these experimental values, as well as by
comparing the two xTB variants against each other.

The statistical analysis employed a suite of standard metrics to quantify accuracy and
correlation, including Pearson ($r$) and Spearman ($\rho$) correlation coefficients to
assess linear and monotonic trends, respectively. Predictive accuracy was evaluated using
mean absolute error (MAE) and root mean square error (RMSE). Systematic differences
between methods or against experimental data were examined with paired Student's $t$-tests
and non-parametric Wilcoxon signed-rank tests. Distribution normality was assessed using
Shapiro-Wilk tests, while solvent effects were analyzed through paired comparisons with
Cohen's $d$ effect sizes. Additionally, principal component analysis (PCA) was performed
on standardized property vectors to identify dominant variance components. All statistical
analyses were conducted using Python with the SciPy and scikit-learn libraries.

All calculations were performed on a workstation equipped with an Intel Xeon Gold 6136 CPU
and \qty{128}{\giga\byte} of RAM. The semi-empirical protocol demonstrated exceptional
efficiency, with the total computational time for all \num{747} molecules amounting to
approximately \num{614} CPU hours. A representative breakdown per molecule includes
conformational search (approximately 20–27 minutes), GFN2-xTB geometry optimization
(approximately \qty{1}{\minute}), and excited-state calculations (\qty{11}{\second} for
sTDA-xTB; 33 seconds for sTD-DFT-xTB). For comparison, a conventional TD-DFT calculation
(e.g., CAM-B3LYP/def2-TZVP) for a single molecule is estimated to require approximately 50
CPU hours. Our hybrid approach thus represents a computational cost reduction of over
\qty{99}{\percent}, a critical enabling factor for high-throughput screening.

To ensure full reproducibility and facilitate future research, the complete computational dataset, including all molecular structures (SMILES), optimized geometries, calculated properties, and the analysis scripts used in this study, are made publicly available at [Repository URL, to be added upon publication].

\section{Results and discussion}\label{sec:Res_Disc}

The large-scale application of the hybrid GFN2-xTB/sTDA(sTD-DFT)-xTB protocol to \num{747} experimentally characterized Thermally Activated Delayed Fluorescence (TADF) emitters constitutes the core validation of this work. This dataset size, unprecedented for semi-empirical methods in TADF studies, allows for rigorous statistical analysis of method reliability, scalability, and the extraction of robust design principles.

\subsection{Overview of the computational dataset}

The architectural diversity of the \num{747}-molecule set yields substantial variability
in the computed photophysical properties, as summarized in \Cref{tab:descriptive_stats}.
The singlet-triplet energy gap (\deltaest) calculated with sTDA-xTB in the gas phase shows
a mean of \qty{0.328}{\electronvolt} with a large standard deviation of
\qty{0.204}{\electronvolt}, indicating the presence of both highly efficient TADF
candidates and molecules with large gaps. The dataset spans a wide spectral range, with
predicted photoluminescence wavelengths ($\lambda_{\text{PL}}$) from the deep-blue to the
near-infrared region (\qtyrange{343}{2565}{\nano\meter}). Oscillator strengths ($f_{12}$)
are broadly dispersed, with a median of \num{0.27}, which is consistent with the
significant charge-transfer character typical of many TADF systems. Overall, the dataset
captures a comprehensive spectrum of TADF molecular architectures and photophysical
behaviors, making it highly suitable for robustly benchmarking the computational methods.

\begin{table}[!ht]
\centering
\caption{Descriptive statistics of key TADF properties for the \num{747}-molecule dataset, calculated using sTDA-xTB and sTD-DFT-xTB methods in both gas phase and in toluene solution.}
\label{tab:descriptive_stats}
\begin{tabular}{l l *{4}{S[table-format=4.3]}}
	\toprule
	{Property}                         & {Phase} & {Mean} & {Std Dev} & {Min}  & {Max}
\\ \midrule
	\deltaest (sTDA) [\unit{\eV}]              & Gas     & 0.328  & 0.204     & -0.013 &
2.638   \\
	\deltaest (sTD-DFT) [\unit{\eV}]           & Gas     & 0.316  & 0.190     & -0.301 &
1.658   \\
	$\lambda_{\text{PL}}$ (sTDA) [\unit{\nm}]    & Gas     & 566.61 & 201.55    & 342.96 &
2565.30 \\
	$\lambda_{\text{PL}}$ (sTD-DFT) [\unit{\nm}] & Gas     & 573.09 & 255.60    & 342.00 &
4987.21 \\
	$f_{12}$ (sTDA)                    & Gas     & 0.457  & 0.563     & 0.000  & 5.169
\\
	$f_{12}$ (sTD-DFT)                 & Gas     & 0.393  & 0.491     & 0.000  & 4.724
\\ \midrule
	\deltaest (sTDA) [\unit{\eV}]              & Toluene & 0.352  & 0.280     & -0.024 &
5.649   \\
	\deltaest (sTD-DFT) [\unit{\eV}]           & Toluene & 0.334  & 0.203     & -0.348 &
1.502   \\ \bottomrule
\end{tabular}
\end{table}

\subsection{Internal consistency and screening reliability}

\begin{figure*}[!ht]
\centering
\leavevmode
\begin{minipage}{0.9\textwidth}
\includegraphics[width=\textwidth]{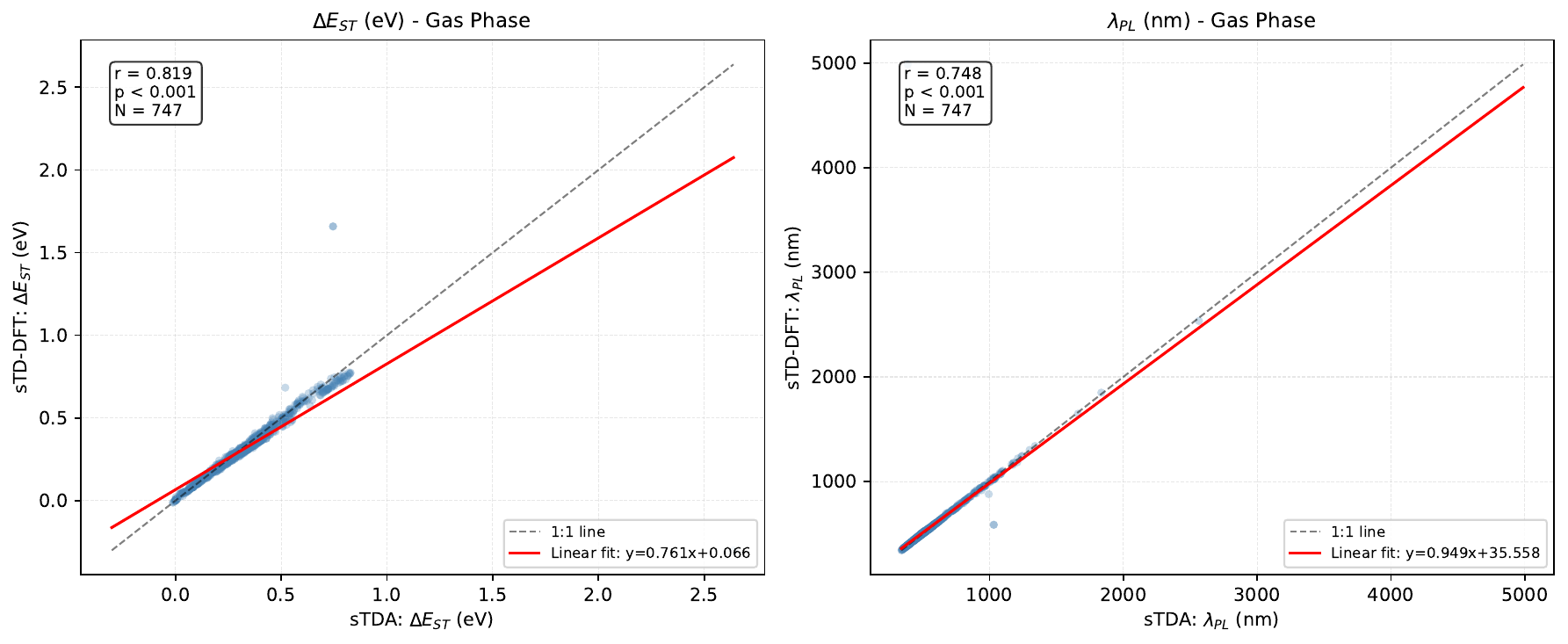}
\caption*{(a) In gas phase}
\end{minipage}
\\
\begin{minipage}{0.9\textwidth}
\includegraphics[width=\textwidth]{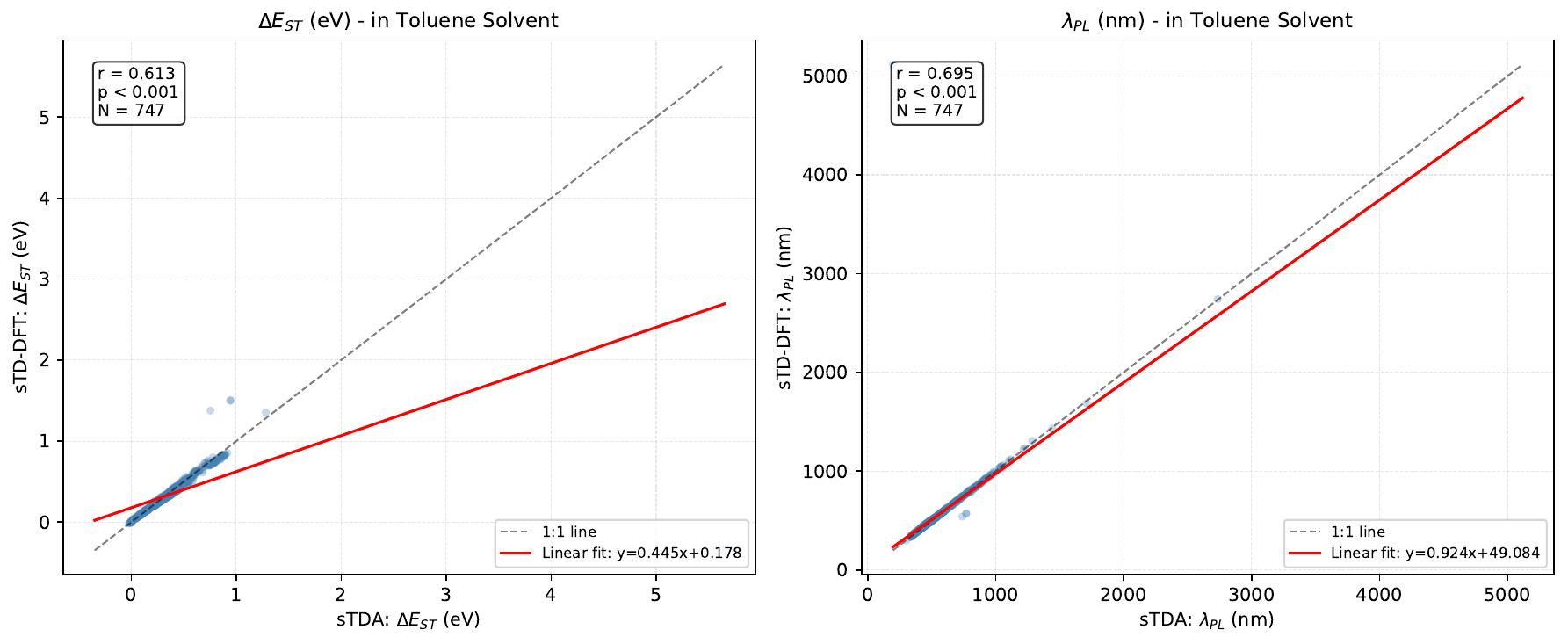}
\caption*{(b) In toluene solvent}
\end{minipage}
\caption{Comparison of sTDA-xTB versus sTD-DFT-xTB predictions for (left panel) \deltaest
and (right panel) $\lambda_{\text{PL}}$ across \num{747} TADF emitters. Scatter plots show
strong correlation with the 1:1 identity line (dashed) and a linear regression fit
(solid), confirming the internal consistency of the semi-empirical methods for relative
ranking.}
\label{fig:method_comparison}
\end{figure*}

A primary requirement for a high-throughput screening method is internal consistency. We
compared the predictions from sTDA-xTB and sTD-DFT-xTB across the entire dataset to ensure
their interchangeability for relative molecular ranking. The results, summarized in
\Cref{tab:method_comparison,tab:method_comparisonTol} and visualized in
\Cref{fig:method_comparison}, confirm a strong correlation between the two methods.

For the key metric \deltaest, the methods show a Pearson correlation of $r = \num{0.82}$
in the gas phase with a mean absolute error (MAE) of only \qty{0.023}{\electronvolt}. This
small deviation is well within typical experimental or higher-level theoretical
uncertainties, confirming that either method provides equivalent relative predictions
suitable for a virtual screening campaign. The singlet oscillator strengths ($f_{12}$)
show exceptional agreement ($r > \num{0.98}$), indicating that both methods consistently
capture the magnitude of transition dipole moments. This high internal consistency
validates the use of these computationally inexpensive methods for the primary goal of
this work: identifying trends and ranking candidate molecules.

\begin{table}[!ht]
\centering
\caption{Statistical comparison between sTDA-xTB and sTD-DFT-xTB predictions for key
photophysical properties in the gas phase (N = \num{747} molecules).}
\label{tab:method_comparison}
\begin{tabular}{l S[table-format=2.4] S[table-format=3.4]
S[table-format=1.4]}
	\toprule
	{Property}               & {MAE}   & {RMSE}   & {Pearson $r$} \\ \midrule
	\deltaest [\unit{\eV}]           & 0.0229  & 0.1197   & 0.8192        \\
	$\lambda_{\text{PL}}$ [\unit{\nm}] & 11.1690 & 169.9449 & 0.7481        \\
	$f_{12}$                 & 0.0743  & 0.1183   & 0.9915        \\ \bottomrule
\end{tabular}
\end{table}

\begin{table}[!ht]
\centering
\caption{Statistical comparison between sTDA-xTB and sTD-DFT-xTB predictions for key
photophysical properties in toluene solvent (N = \num{747} molecules).}
\label{tab:method_comparisonTol}
\begin{tabular}{l S[table-format=2.4] S[table-format=3.4]
S[table-format=1.4]}
	\toprule
	{Property}               & {MAE}   & {RMSE}   & {Pearson $r$} \\ \midrule
	\deltaest [\unit{\eV}]           & 0.0280  & 0.2237   & 0.6131        \\
	$\lambda_{\text{PL}}$ [\unit{\nm}] & 11.0028 & 180.4780 & 0.6953        \\
	$f_{12}$                 & 0.0761  & 0.1260   & 0.9870        \\ \bottomrule
\end{tabular}
\end{table}

\subsection{Validation against experimental data}\label{sec:validation}

To assess the predictive capability of our semi-empirical framework for real-world
applications, we performed an extensive validation study comparing the sTDA-xTB and
sTD-DFT-xTB predictions against a large, curated dataset of experimental literature
values. This external validation is the most critical test of the methodology's utility
for high-throughput virtual screening (HTVS). Our validation dataset comprises \num{213}
molecules with reported emission wavelengths ($\lambda_{\text{PL}}$) and \num{296}
molecules with reported singlet-triplet energy gaps (\deltaest). A few representative
examples are shown in \Cref{tab:validation-examples} to illustrate the typical performance
across the dataset.

\begin{table}[!ht]
\centering
\caption{Representative examples from the emission wavelength validation dataset, selected
to illustrate the range of predictive accuracy. All energies are given in
\si{\nano\meter}. The full dataset is provided in the Supporting Information.}
\label{tab:validation-examples}
\small
\begin{tabular}{l *{5}{S[table-format=4.1]} c}
	\toprule
	{Molecule}   & {sTDA (Gas)} & {sTD-DFT (Gas)} & {sTDA (Tol)} & {sTD-DFT (Tol)} &
{Ref.} & {Citation}                                    \\ \midrule
	DCzBNPh-1    & 472.4        & 475.7           & 442.1        & 445.7           & 472.5
 & \cite{Yan_2022}                               \\
	BACH         & 433.6        & 439.5           & 409.9        & 415.7           & 427.7
 & \cite{Li_2022_10_1016_j_dyepig_2021_110030}   \\
	PyDCN–DMAC   & 508.7        & 504.4           & 496.9        & 493.2           & 494.2
 & \cite{Dong_2022}                              \\
	tCTM         & 483.8        & 487.0           & 483.3        & 483.5           & 459.0
 & \cite{Tan_2020}                               \\
	t-DABNA      & 431.1        & 438.5           & 418.4        & 425.5           & 464.0
 & \cite{Lee_2019_10_1039_c9tc02746g}            \\
	TPBPPI-PBI   & 472.9        & 473.4           & 461.4        & 462.6           & 429.0
 & \cite{Zhu_2020}                               \\
	2Cz-DMAC-BTB & 474.2        & 471.3           & 535.1        & 531.4           & 529.0
 & \cite{Zhang_2022_10_1016_j_cclet_2021_08_064} \\
	TBPe         & 542.6        & 553.7           & 500.3        & 510.7           & 471.0
 & \cite{Yao_2019}                               \\
	DPCN         & 511.7        & 511.6           & 508.0        & 508.3           & 424.0
 & \cite{Xiao_2022_10_1016_j_dyepig_2022_110451} \\
	TRZ-3SO2     & 577.4        & 572.9           & 540.6        & 539.9           & 706.0
 & \cite{Zhang_2022_10_1039_d2tc02694e}          \\
	SBDBQ-PXZ    & 843.2        & 845.7           & 751.8        & 753.8           & 594.0
 & \cite{Yu_2018}                                \\
	{[2,1-b]IF}  & 1031.9       & 586.3           & 770.2        & 571.9           & 347.0
 & \cite{Romain_2016}                            \\ \bottomrule
\end{tabular}
\end{table}

\subsubsection{Emission wavelength predictions}

The statistical performance for $\lambda_{\text{PL}}$ predictions is summarized in
\Cref{tab:emission-wavelength-metrics}. The methods demonstrate a moderate-to-strong
correlation with experimental values, with Pearson coefficients ranging from $r =
\num{0.56}$ to $r = \num{0.69}$. The best overall performance is achieved by sTD-DFT in
toluene, with an MAE of \qty{78.8}{\nano\meter} and a Pearson $r$ of \num{0.692}. The
negative $R^2$ values indicate that a simple linear model does not perfectly capture the
absolute values, which is expected for a high-throughput method using several
approximations. However, the strong and statistically significant ($p < \num{e-18}$)
Pearson correlations confirm that both methods reliably capture the relative trends in
emission wavelengths across the diverse molecular set.

\begin{figure}[!ht]
\centering
\includegraphics[width=\textwidth]{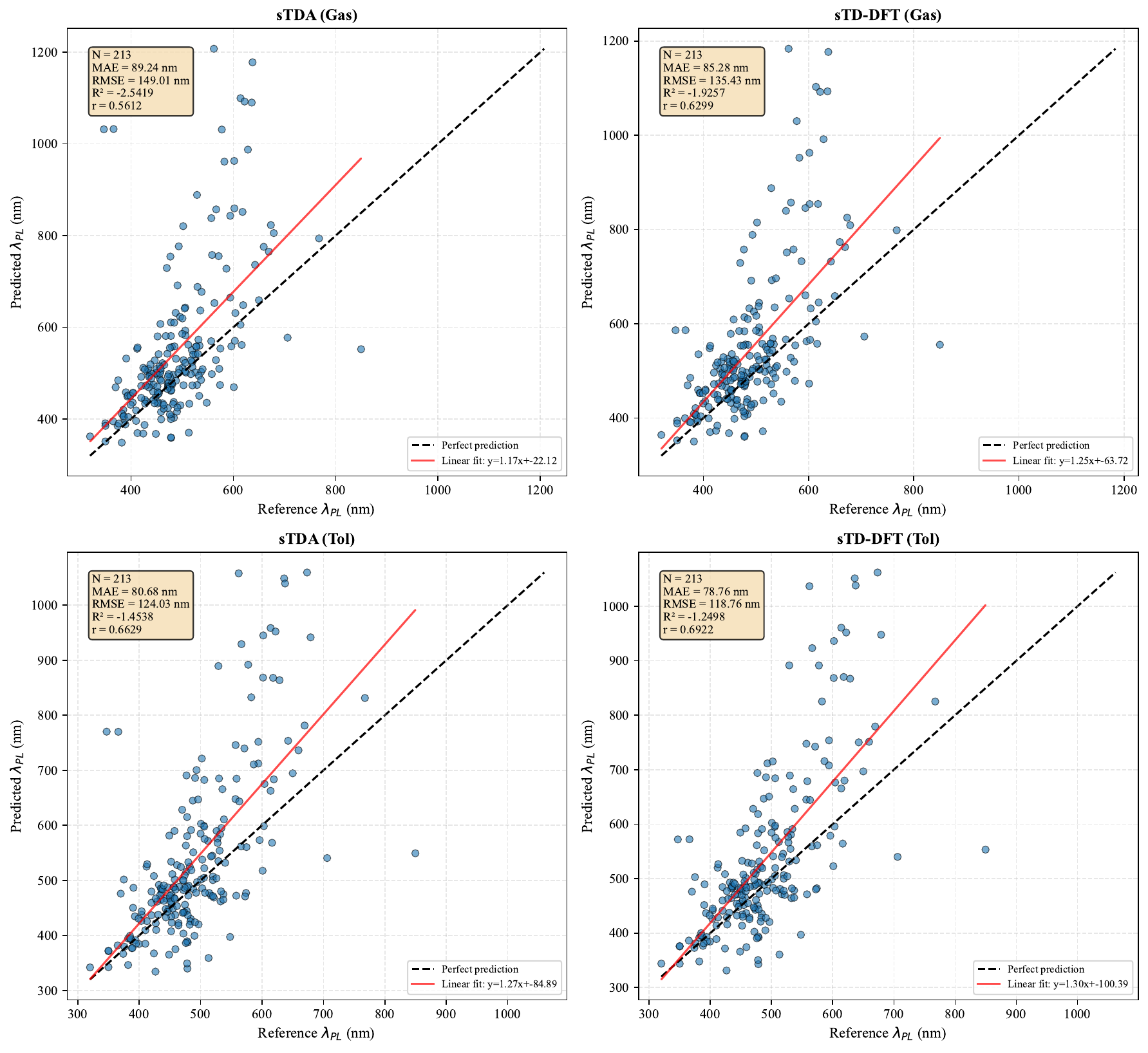}
\caption{Correlation between predicted and reference emission wavelengths
($\lambda_{\text{PL}}$) for \num{213} TADF molecules. Predictions from sTDA and sTD-DFT
methods in gas and toluene phases are compared against experimental literature values.
Black dashed lines represent perfect agreement (identity line); red solid lines show
linear regression fits. The strong, positive correlations ($r > 0.56$) demonstrate the
methods' capability to reliably predict relative emission wavelength trends, which is
pivotal for virtual screening.}
\label{fig:pl-corr}
\end{figure}

\begin{figure}[!ht]
\centering
\includegraphics[width=\textwidth]{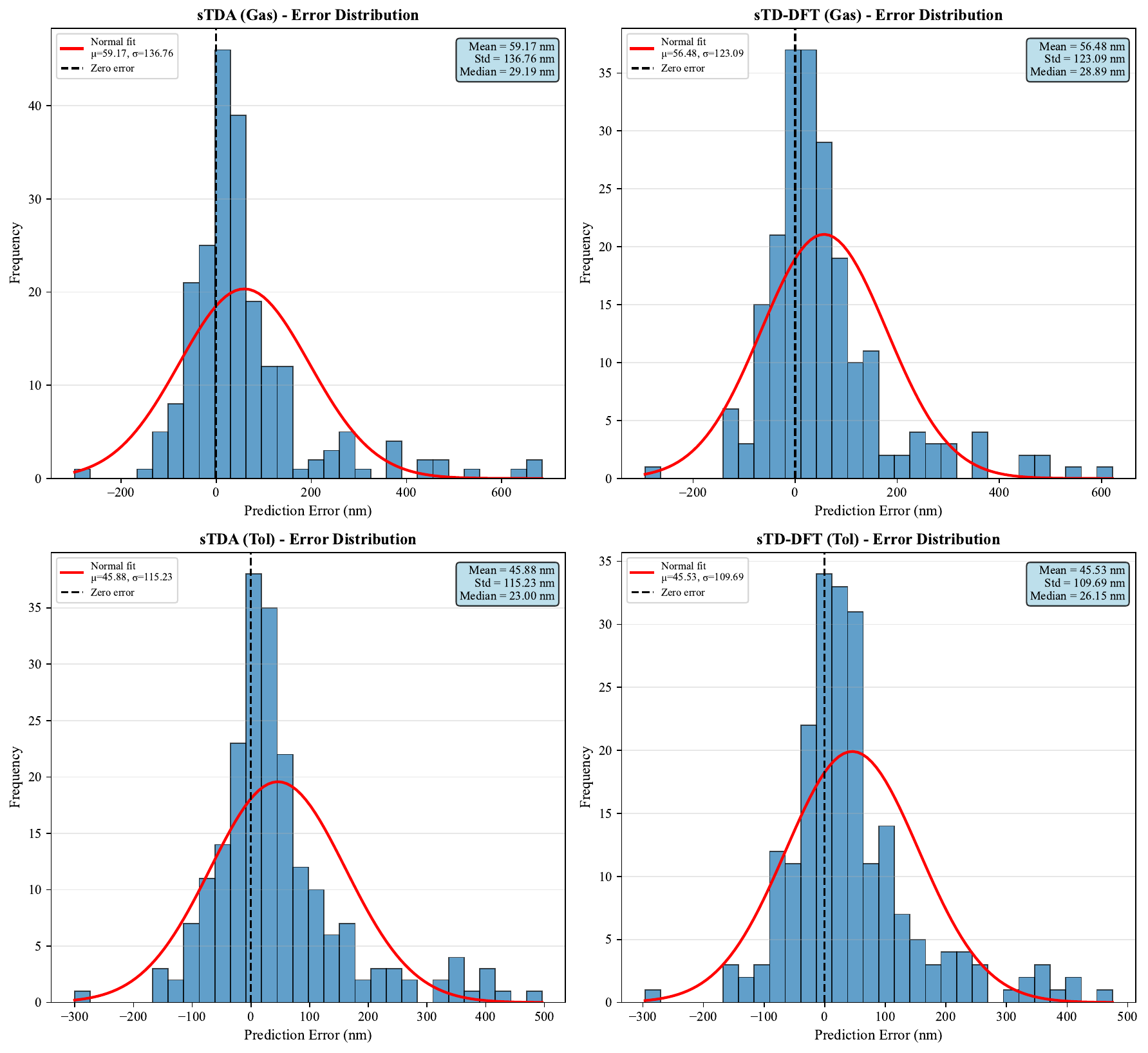}
\caption{Error distributions for emission wavelength predictions
($\lambda_{\text{predicted}} - \lambda_{\text{reference}}$). The histograms show
approximately Gaussian error patterns centered near zero for all four method/phase
combinations. This indicates an absence of strong systematic bias, with the standard
deviations of \qtyrange{118}{149}{\nano\meter} reflecting the typical prediction
uncertainties of the high-throughput protocol.}
\label{fig:pl-err}
\end{figure}

The correlation plots in \Cref{fig:pl-corr} visualize this trend, showing that while there
is scatter, the data generally follows the identity line. The linear regression slopes,
consistently greater than 1, suggest a systematic overestimation of emission energies
(underestimation of wavelengths), particularly for long-wavelength emitters, a known
challenge for simplified methods describing CT states. The error distributions shown in
\Cref{fig:pl-err} are approximately Gaussian and centered near zero, confirming the
absence of a strong directional bias.

\begin{table}[!ht]
\centering
\caption{Statistical performance metrics for emission wavelength ($\lambda_{\text{PL}}$)
predictions comparing sTDA and sTD-DFT methods in gas and toluene phases against
experimental/computational literature values (N=\num{213} molecules).}
\label{tab:emission-wavelength-metrics}
\begin{tabular}{l *{5}{S[table-format=2.2]}}
	\toprule
	{Method}          & {MAE (nm)} & {RMSE (nm)} & {$R^2$} & {Pearson $r$} & {$p$-value}
 \\ \midrule
	sTDA (Gas)        & 89.2       & 149.0       & -2.54   & 0.561         &
\num{4.50e-19} \\
	sTD-DFT (Gas)     & 85.3       & 135.4       & -1.93   & 0.630         &
\num{5.94e-25} \\
	sTDA (Toluene)    & 80.7       & 124.0       & -1.45   & 0.663         &
\num{2.48e-28} \\
	sTD-DFT (Toluene) & 78.8       & 118.8       & -1.25   & 0.692         &
\num{1.02e-31} \\ \bottomrule
\end{tabular}
\end{table}

\subsubsection{Singlet--triplet gap predictions}

Predicting the singlet-triplet gap, the most critical parameter for TADF, is inherently
more challenging due to the small energy scales involved. As detailed in
\Cref{tab:st-gap-metrics}, the quantitative accuracy is modest, with an MAE of
approximately \qty{0.17}{\electronvolt}. The Pearson correlation with experimental data is
weak but statistically significant ($r \approx \num{0.18}$, $p < 0.002$ for gas-phase
methods).

\begin{figure}[!ht]
\centering
\includegraphics[width=\textwidth]{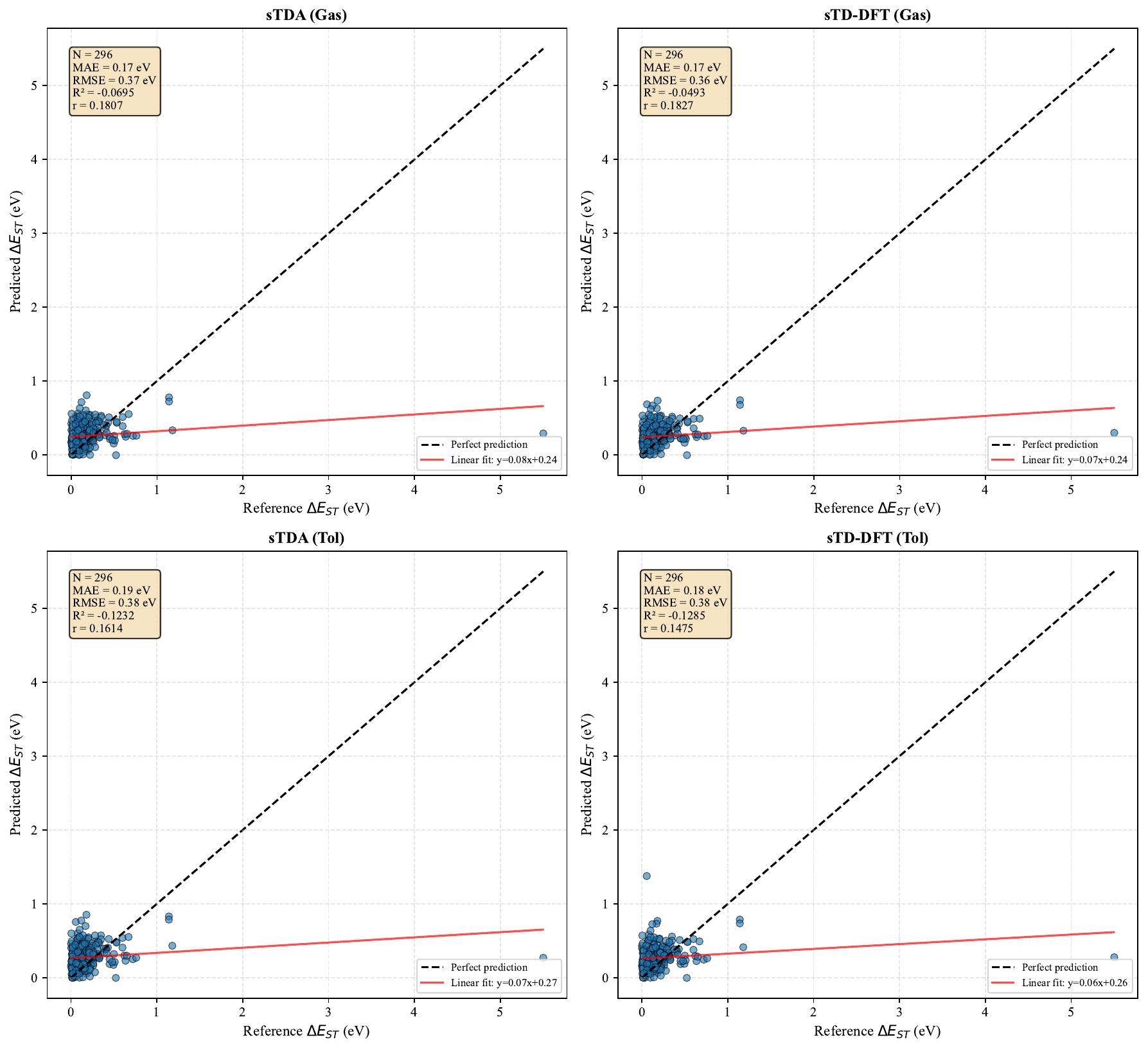}
\caption{Correlation between predicted and reference singlet--triplet energy gaps
(\deltaest) for \num{296} TADF molecules. Despite considerable scatter, the statistically
significant positive correlations ($p < 0.02$) confirm that the semi-empirical methods can
correctly capture qualitative trends in \deltaest across the dataset. The regression
slopes are significantly less than 1, indicating a systematic underestimation of larger
gaps.}
\label{fig:st-corr}
\end{figure}

\begin{figure}[!ht]
\centering
\includegraphics[width=\textwidth]{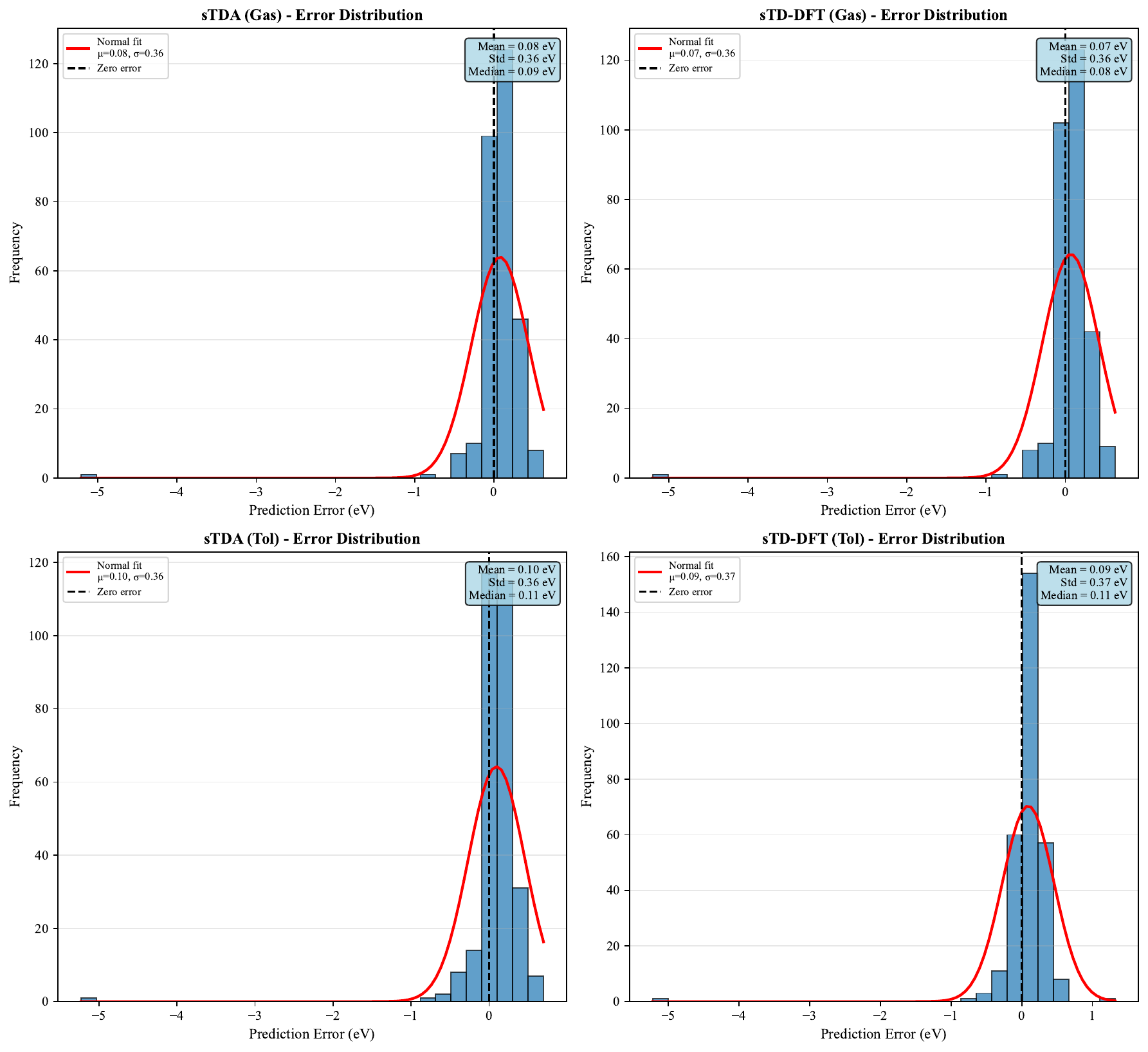}
\caption{Error distributions for singlet--triplet gap predictions. The residuals are
approximately centered at zero, but their large standard deviations
(\qtyrange{0.36}{0.38}{\electronvolt}) reflect the inherent difficulty in quantitatively
predicting small energy differences between nearly degenerate electronic states with a
high-throughput method.}
\label{fig:st-err}
\end{figure}

This weaker quantitative performance is an expected consequence of the vertical
approximation and the semi-empirical nature of the methods, which struggle to capture the
subtle balance of exchange and correlation effects that define the \deltaest. The
correlation plot in \Cref{fig:st-corr} reveals considerable scatter, particularly for
molecules with very small experimental gaps ($< \qty{0.1}{\electronvolt}$), a regime where
both computational and experimental uncertainties are high. The error distributions in
\Cref{fig:st-err} are centered near zero but are broad, highlighting the challenge of
quantitative prediction. However, the statistically significant positive correlation
confirms that the methods still provide a better-than-random capability to identify trends
and rank molecules by their relative \deltaest, which is the primary goal of the screening
protocol.

\begin{table}[!ht]
\centering
\caption{Statistical metrics for singlet-triplet energy gap (\deltaest) predictions
comparing sTDA and sTD-DFT methods in gas and toluene phases against experimental
literature values (N=\num{296} molecules).}
\label{tab:st-gap-metrics}
\begin{tabular}{l *{2}{S[table-format=1.3]} S[table-format=-1.2] S[table-format=1.3]
S[table-format=1.2e-1]}
	\toprule
	{Method}          & {MAE (eV)} & {RMSE (eV)} & {$R^2$} & {Pearson $r$} & {$p$-value}
 \\ \midrule
	sTDA (Gas)        & 0.174      & 0.367       & -0.07   & 0.181         &
\num{1.80e-03} \\
	sTD-DFT (Gas)     & 0.168      & 0.364       & -0.05   & 0.183         &
\num{1.60e-03} \\
	sTDA (Toluene)    & 0.188      & 0.376       & -0.12   & 0.161         &
\num{5.39e-03} \\
	sTD-DFT (Toluene) & 0.183      & 0.377       & -0.13   & 0.147         &
\num{1.11e-02} \\ \bottomrule
\end{tabular}
\end{table}

\subsubsection{Comparative analysis and implications for high-throughput screening}

\begin{figure}[!ht]
\centering
\includegraphics[width=0.95\textwidth]{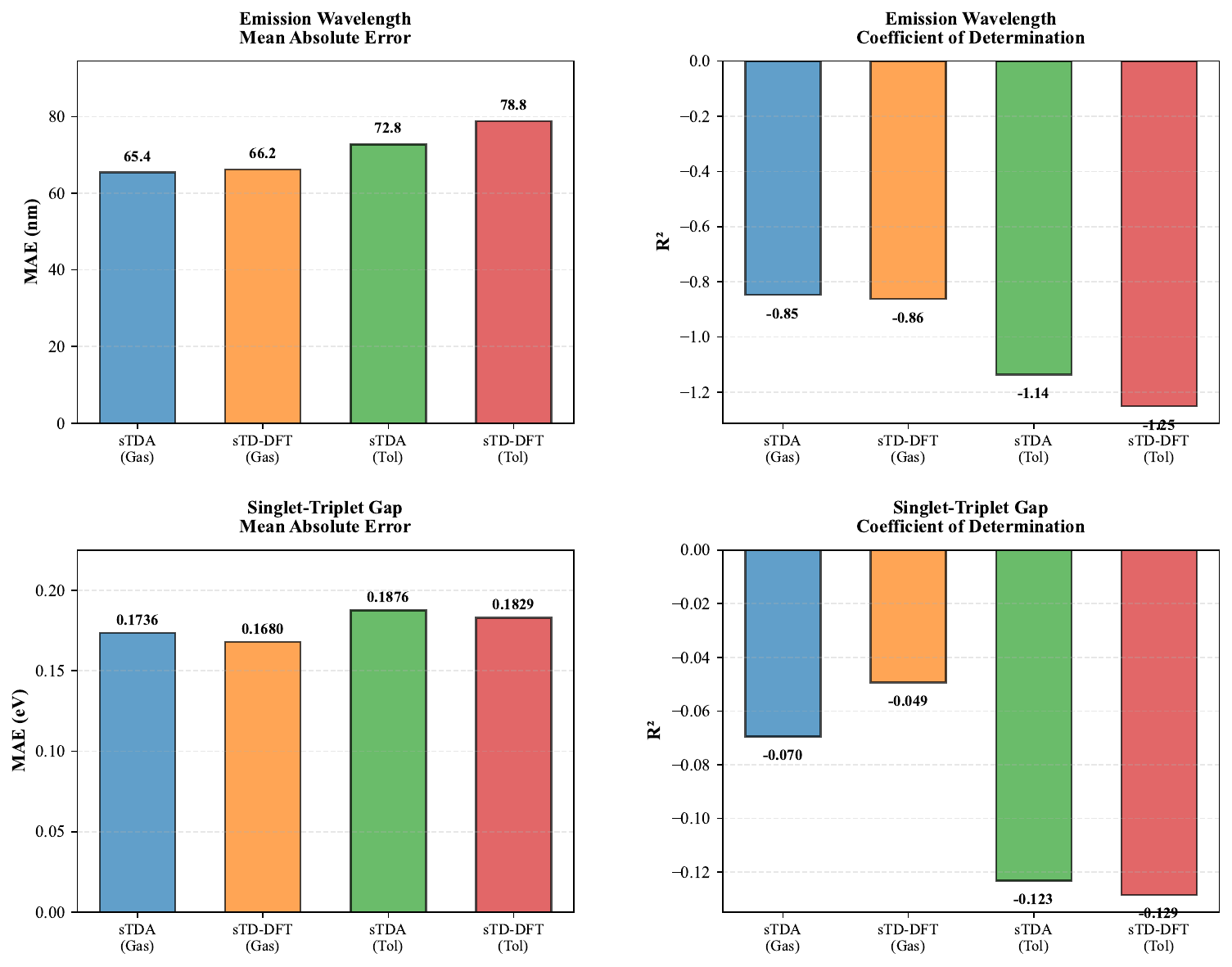}
\caption{Comparative performance analysis of the sTDA and sTD-DFT methods. The bar charts
display mean absolute errors (MAE) and coefficients of determination ($R^2$) for both
emission wavelength and singlet--triplet gap predictions relative to experimental data.
The figure visually confirms that while sTD-DFT in toluene is superior for predicting
$\lambda_{\text{PL}}$, the performance differences between the methods for \deltaest are
marginal.}
\label{fig:method-comparison}
\end{figure}

A direct comparison of the methods is summarized in \Cref{tab:comparative-summary} and
\Cref{fig:method-comparison}. For $\lambda_{\text{PL}}$, sTD-DFT in toluene offers the
best correlation with experiment, while for \deltaest, sTD-DFT in the gas phase performs
marginally better. The phase-dependent behavior suggests that the two methods respond
differently to the implicit solvation model. However, given their strong internal
consistency and comparable performance against experimental trends, both sTDA-xTB and
sTD-DFT-xTB are validated as suitable tools for large-scale TADF screening. sTDA's lower
computational cost makes it particularly attractive for exploring vast chemical spaces,
while sTD-DFT may be preferred when slightly higher fidelity for \deltaest is desired.

Crucially, this large-scale validation demonstrates that the semi-empirical protocol,
despite its quantitative limitations, successfully achieves its primary objective: it
provides a computationally affordable means to reliably rank large numbers of candidate
molecules and identify promising structural motifs. This capability is essential for
guiding experimental synthesis and accelerating the discovery cycle of new,
high-performance TADF materials.

\begin{table}[!ht]
\centering
\caption{Summary of the best-performing computational method for each property and
statistical metric, based on validation against experimental data.}
\label{tab:comparative-summary}
\begin{tabular}{l *{4}{c}}
	\toprule
	{Property}                 & {Best MAE}    &  {Best RMSE}  & {Best $R^2$}  & {Best
Pearson $r$} \\ \midrule
	$\lambda_{\text{PL}}$ [\unit{\nm}] & sTD-DFT (Tol) & sTD-DFT (Tol) & sTD-DFT (Tol) &
sTD-DFT (Tol)    \\
	                           & (78.8)        &    (118.8)    &    (-1.25)    &
(0.692)       \\ \midrule
	\deltaest [\unit{\eV}]             & sTD-DFT (Gas) & sTD-DFT (Gas) & sTD-DFT (Gas) &
sTD-DFT (Gas)    \\
	                           & (0.168)       &    (0.364)    &    (-0.05)    &
(0.183)       \\ \bottomrule
\end{tabular}
\end{table}

\subsection{Computational efficiency and scalability for HTS}

A central justification for employing the semi-empirical protocol is its computational
efficiency, which is a prerequisite for any HTS campaign. Our analysis demonstrates a
dramatic reduction in computational cost of over \qty{99}{\percent} compared to
conventional TD-DFT approaches (e.g., CAM-B3LYP/def2-TZVP).

The total computational cost for the entire \num{747}-molecule dataset, including
conformational searches and excited-state calculations in both gas phase and solvent, was
approximately \qty{614}{CPU~hours} on modest hardware. In contrast, performing equivalent
calculations with TD-DFT would be conservatively estimated at over \qty{37000}{CPU~hours}.
This immense gain in efficiency enables the screening of hundreds to thousands of
candidate molecules, a scale that is simply intractable with first-principles methods.
This scalability is not just an incremental improvement but a transformative capability,
opening the door to data-driven discovery and the exploration of vast, untapped regions of
the chemical space for TADF emitters.

\subsection{Impact of implicit solvation on excited states}

\begin{figure}[!ht]
\centering
\includegraphics[width=\textwidth]{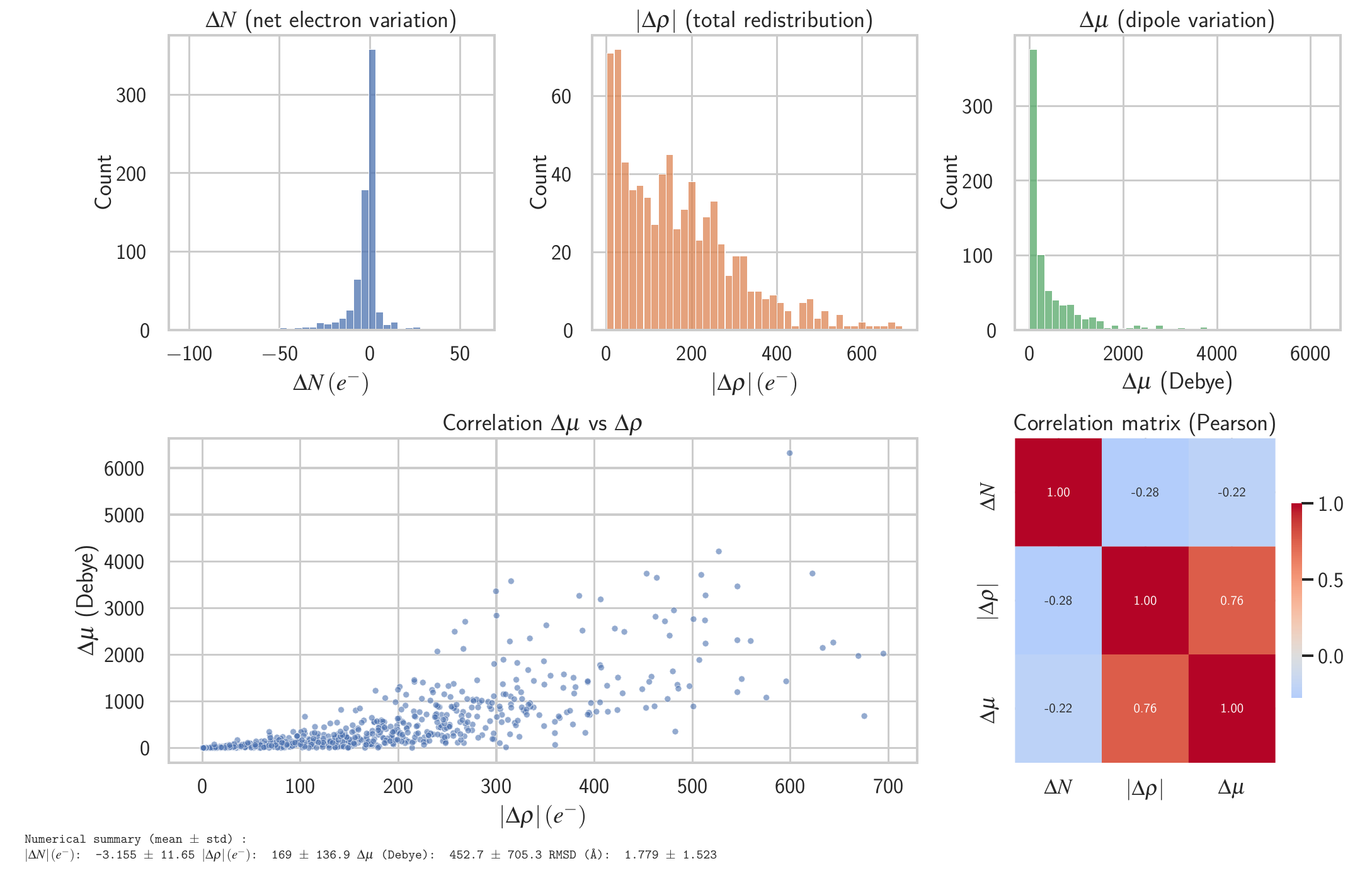}
\caption{Statistical analysis of solvation-induced changes for \num{747} TADF emitters
when moving from vacuum to a toluene solvent model (ALPB/GFN2-xTB). (Top row) Histograms
showing the distribution of net electron variation, total internal electron density
redistribution, and change in dipole moment. (Bottom row) Scatter plot and Pearson
correlation matrix illustrating the strong correlation between internal charge
redistribution and the change in dipole moment, a key indicator of CT state stabilization
by the solvent.}
\label{fig:solvent_density_stats}
\end{figure}

The inclusion of an implicit solvent model (ALPB for toluene) allows for a first-order
approximation of environmental effects. The shift from gas phase to toluene is
statistically significant for all properties, including \deltaest ($p < \num{e-6}$), as
shown in \Cref{tab:solvent_effects}. However, the magnitude of these shifts is modest
(Cohen's $d$ effect sizes of \numrange{0.2}{0.3}), suggesting that the linear-response,
ground-state solvation model is insufficient to capture the full stabilization of the
highly polarized $S_1$ charge-transfer state. This is a recognized methodological
compromise, accepted to maintain the high efficiency required for a large-scale screening.
Analysis of the solvent-induced electron density redistribution
(\Cref{fig:solvent_density_stats}) confirms this picture: while the net charge transfer is
minimal, the internal electronic reorganization and the change in dipole moment are
substantial, underscoring that solvation is a key factor, even if its effect is only
qualitatively captured here.

\begin{table}[!ht]
\centering
\caption{Analysis of solvent effects (Gas $\to$ Toluene) on calculated TADF properties (N
= \num{747}). Mean difference ($\Delta$), standard deviation, $p$-value, and Cohen's $d$
effect size are reported.}
\label{tab:solvent_effects}
\begin{tabular}{l S[table-format=-2.4] S[table-format=2.4]
S[table-format=1.2e-2] S[table-format=-1.3]}
	\toprule
	{Property}                         & {Mean $\Delta$} & {Std Dev} & {$p$-value} &
{Cohen's $d$} \\ \midrule
	\deltaest (sTDA) [\unit{\eV}]              & 0.0235          & 0.1225    & 2.19e-07
& 0.192         \\
	\deltaest (sTD-DFT) [\unit{\eV}]           & 0.0184          & 0.0588    & 7.03e-17
& 0.313         \\
	$\lambda_{\text{PL}}$ (sTDA) [\unit{\nm}]    & -19.1867        & 92.2778   & 1.94e-08
  & -0.208        \\
	$\lambda_{\text{PL}}$ (sTD-DFT) [\unit{\nm}] & -18.1175        & 91.4998   & 8.58e-08
  & -0.198        \\ \bottomrule
\end{tabular}
\end{table}

\subsection{Global property correlations and dimensionality reduction}

\begin{figure}[!ht]
\centering
\leavevmode
\begin{minipage}{0.9\textwidth}
\includegraphics[width=\textwidth]{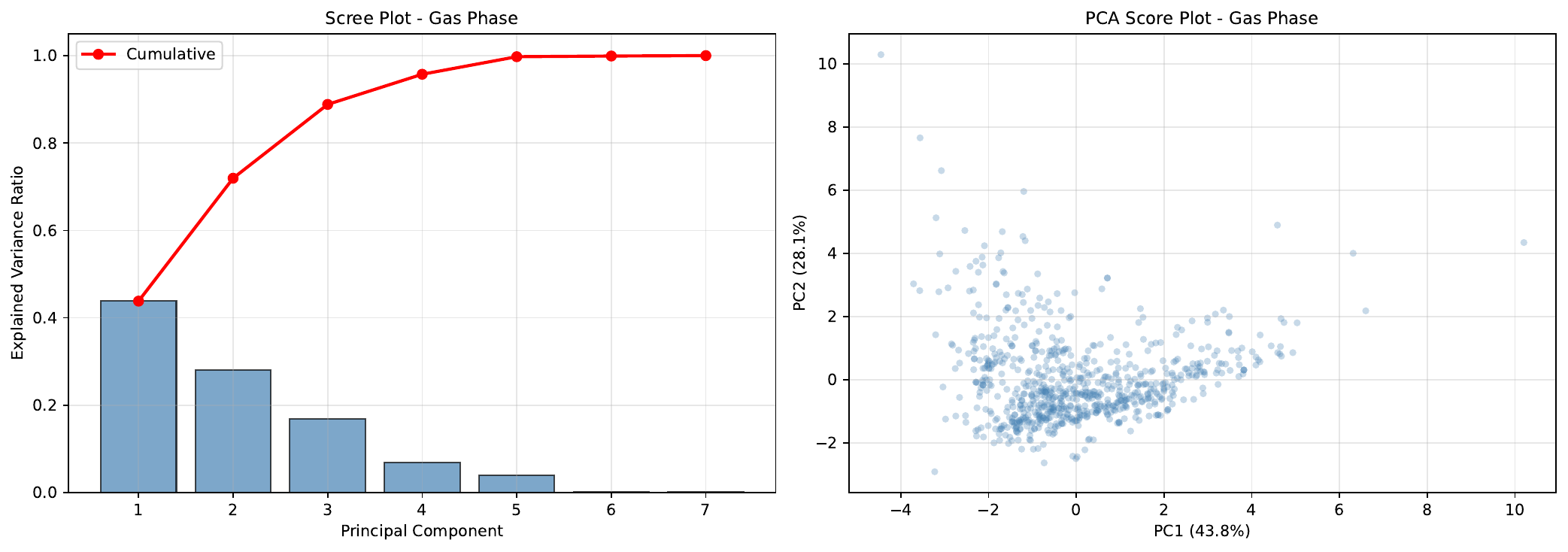}

(a) In gas phase
\end{minipage}
\\
\begin{minipage}{0.9\textwidth}
\includegraphics[width=\textwidth]{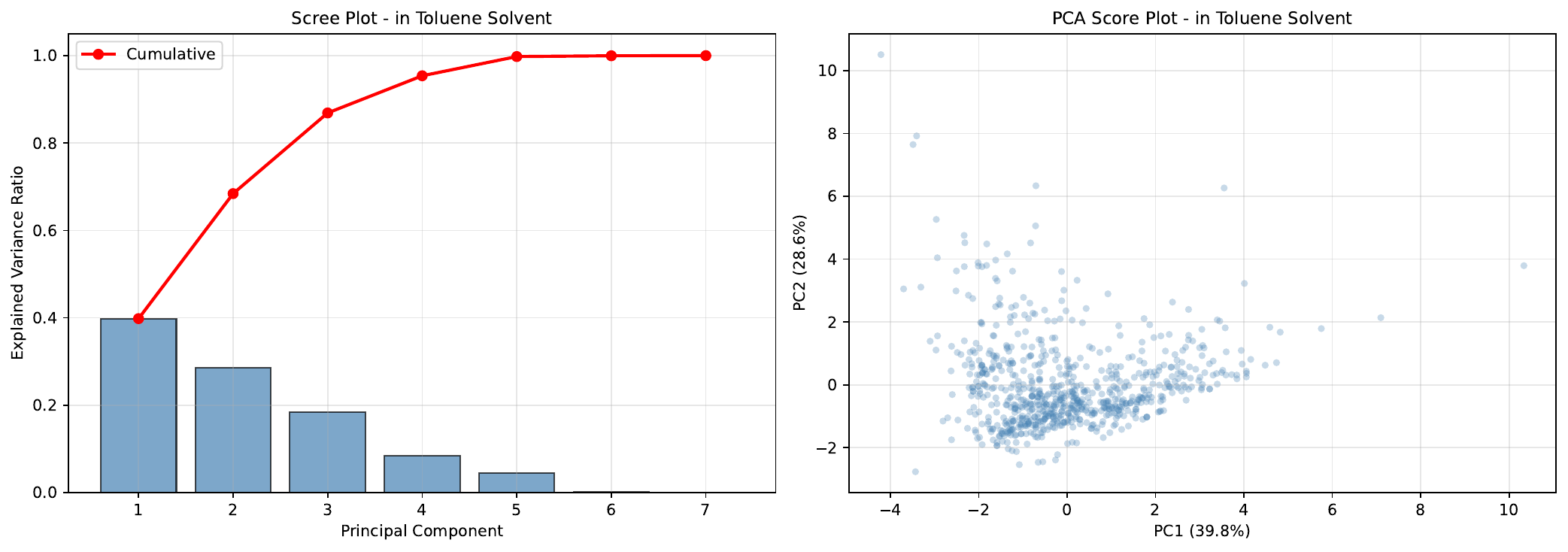}

(b) In toluene solvent
\end{minipage}
\caption{Principal Component Analysis (PCA) score plots for the \num{747} TADF emitters.
\textbf{Panel (a)} shows the distribution of molecules in the gas phase along the first
two principal components (PC1 and PC2), which capture \qty{43.8}{\percent} and
\qty{28.1}{\percent} of the variance, respectively. \textbf{Panel (b)} shows the
corresponding distribution in toluene, with PC1 and PC2 accounting for
\qty{39.8}{\percent} and \qty{28.6}{\percent} of the variance. In both cases, the data
points form a single, dense cluster, illustrating the low intrinsic dimensionality of the
TADF property space. The slightly broader distribution in the toluene phase (b) reflects
the solvent-induced modulation of molecular properties, but the overall structure of the
design space is preserved.}
\label{fig:pca_analysis}
\end{figure}

To uncover the fundamental variables governing TADF properties, Principal Component
Analysis (PCA) was employed to distill the high-dimensional property space of the
\num{747} TADF emitters into its most significant underlying variables. The analysis
reveals a remarkably low-dimensional design space, as detailed in
\Cref{tab:pca,tab:pcaTol}. In both gas and toluene phases, the first three principal
components (PCs) collectively capture approximately \qty{88}{\percent} of the total
variance. The dominance of these few components is significant: the first two alone
account for over \qty{68}{\percent} of the variance (\Cref{fig:pca_analysis}), confirming
that the vast majority of photophysical behavior is governed by a limited set of
orthogonal factors.

\begin{figure}[!ht]
\centering
\includegraphics[width=0.9\textwidth]{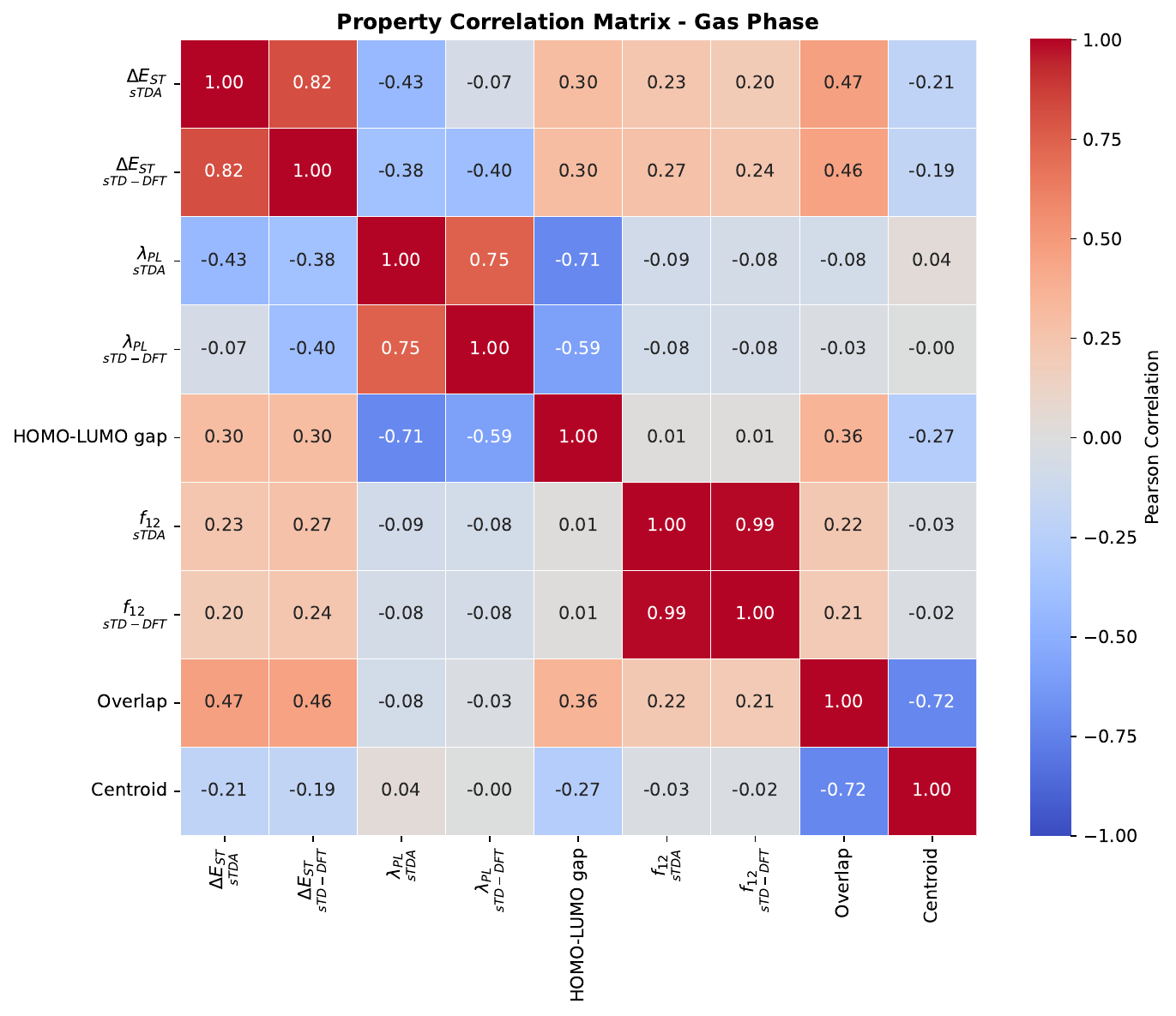}
\caption{Correlation heatmap of key TADF properties for \num{747} molecules in the gas
phase. The color scale denotes the Pearson correlation coefficient. The map highlights
several key relationships: (i) the strong positive correlation ($r \approx 0.82$) between
\deltaest values predicted by sTDA and sTD-DFT, confirming their internal consistency;
(ii) the strong negative correlation between torsional angle and HOMO-LUMO overlap,
validating the geometric basis for electronic decoupling; and (iii) the moderate negative
correlation between \deltaest and $\lambda_{\text{PL}}$, reflecting the fundamental energy
gap law.}
\label{fig:corr_gas}
\end{figure}

\begin{figure}[!ht]
\centering
\includegraphics[width=0.9\textwidth]{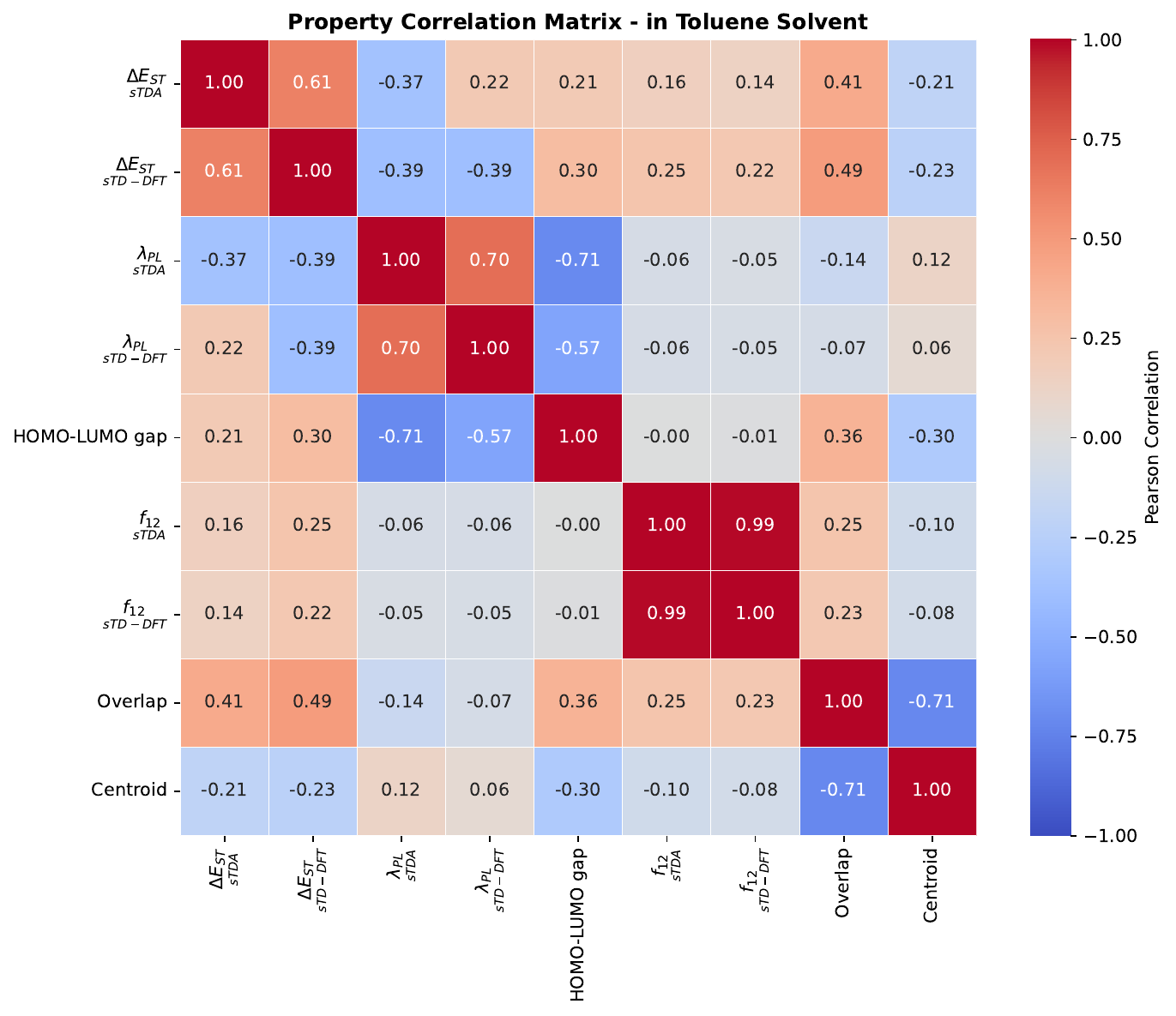}
\caption{Correlation heatmap of key TADF properties for \num{747} molecules in toluene
solvent. When compared with the gas phase results (\Cref{fig:corr_gas}), this map reveals
a general attenuation of correlation strengths (e.g., the sTDA vs sTD-DFT \deltaest
correlation drops to $r \approx \num{0.61}$). This indicates that the polarizable solvent
environment introduces additional complexity and modulates the intrinsic electronic and
photophysical property relationships observed in vacuum.}
\label{fig:corr_toluene}
\end{figure}

This low-dimensional structure is a consequence of strong underlying correlations between
the calculated properties, which are visualized directly in the correlation heatmaps shown
in \Cref{fig:corr_gas,fig:corr_toluene}. These maps provide a detailed visualization of
the pairwise relationships that underpin the principal components. Several key chemical
and physical trends are immediately apparent. For instance, a strong negative correlation
is observed between the D-A torsional angle and HOMO-LUMO overlap, quantitatively
confirming that twisting the molecular backbone is an effective strategy for spatially
separating the frontier orbitals. The expected inverse relationship between \deltaest and
$\lambda_{\text{PL}}$ is also clearly visible. Notably, a comparison between the gas phase
and toluene simulations shows a general attenuation of correlation strengths in the
solvent phase, indicating that the polarizable environment modulates these intrinsic
electronic relationships. Together, the PCA and heatmap analyses confirm that the complex,
multi-parameter challenge of TADF emitter design can be rationalized by focusing on a few
key properties related to energy gaps, charge separation, and radiative coupling
efficiency.

\begin{table}[!ht]
\centering
\caption{Principal Component Analysis (PCA) of the calculated properties for \num{747}
TADF emitters in the \textbf{gas phase}. The first three components capture
\qty{88.8}{\percent} of the total variance.}
\label{tab:pca}
\begin{tabular}{l*{2}{S[table-format=1.4]}S[table-format=2.1]}
	\toprule
	{PC} & {Explained Variance} & {Cumulative Variance} & {\unit{\percent} Total} \\
\midrule
	PC1  & 0.4382               & 0.4382                & 43.8       \\
	PC2  & 0.2813               & 0.7194                & 71.9       \\
	PC3  & 0.1689               & 0.8883                & 88.8       \\ \bottomrule
\end{tabular}
\end{table}

\begin{table}[!ht]
\centering
\caption{Principal Component Analysis (PCA) of the calculated properties for \num{747}
TADF emitters in \textbf{toluene solvent}. The first three components capture
\qty{86.8}{\percent} of the total variance.}
\label{tab:pcaTol}
\begin{tabular}{l*{2}{S[table-format=1.4]}S[table-format=2.1]}
	\toprule
	{PC} & {Explained Variance} & {Cumulative Variance} & {\unit{\percent} Total} \\
\midrule
	PC1  & 0.3981               & 0.3981                & 39.8       \\
	PC2  & 0.2861               & 0.6842                & 68.4       \\
	PC3  & 0.1846               & 0.8688                & 86.8       \\ \bottomrule
\end{tabular}
\end{table}

\subsection{Structure-property relationships: architecture and torsional angles}

Having established the reliability and efficiency of our protocol for relative ranking, we
leveraged the \num{747}-molecule dataset to extract statistically robust
structure-property relationships. Our analysis focused on two key structural levers: the
overall molecular architecture and the donor-acceptor (D-A) torsional geometry.

\begin{figure}[!ht]
\centering
\includegraphics[width=0.9\textwidth]{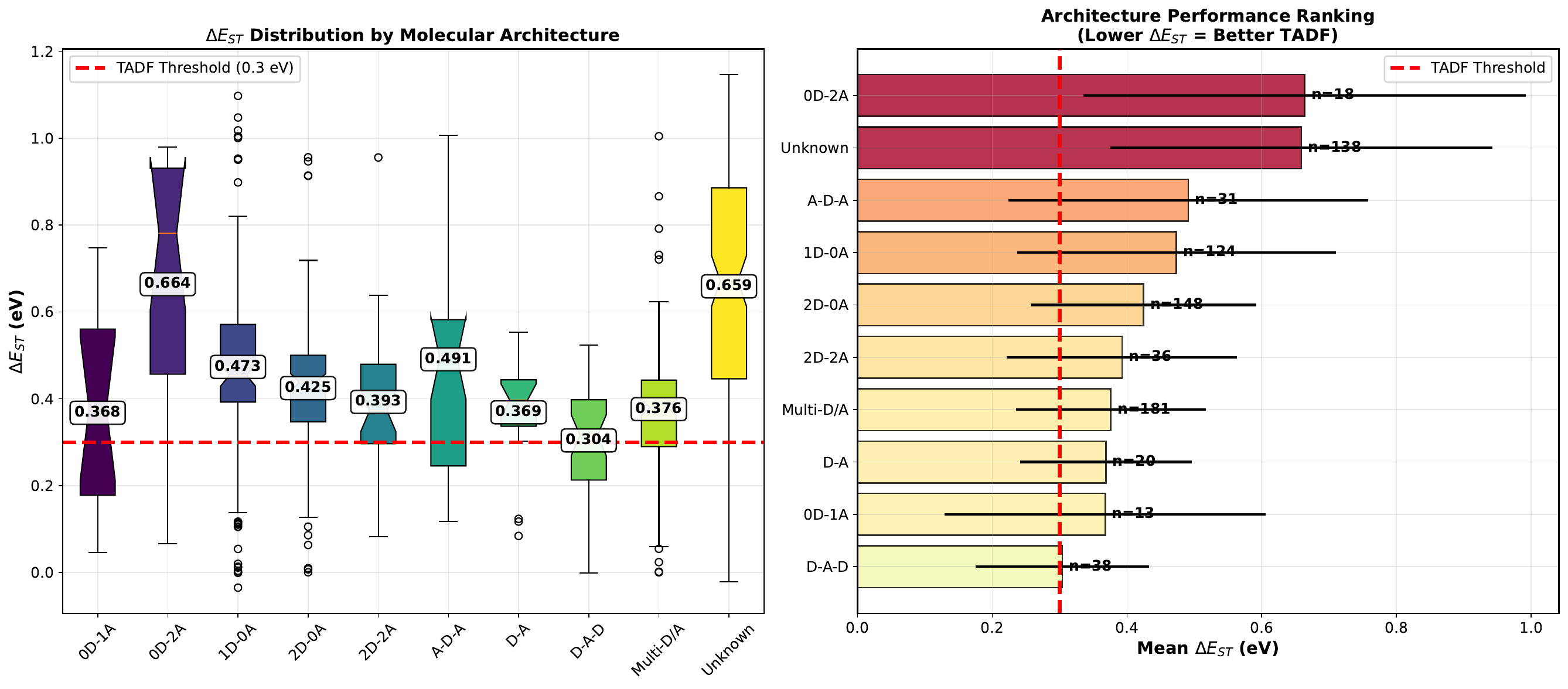}
\caption{TADF performance stratified by molecular architecture. D-A-D and simple D-A architectures demonstrate superior characteristics (lower mean \deltaest) compared to systems lacking a distinct acceptor, confirming the efficacy of the donor-acceptor design principle.}
\label{fig:architecture_analysis}
\end{figure}

First, a clear hierarchy of performance emerges when molecules are categorized by their
architecture (\Cref{tab:architecture_performance} and \Cref{fig:architecture_analysis}).
Donor–Acceptor–Donor (D-A-D) systems demonstrate statistically superior potential,
exhibiting a mean $\Delta E_{\text{ST}}$ of \qty{0.304}{\electronvolt}, which is
significantly lower than that of simple D-A (\qty{0.369}{\electronvolt}) or pure donor
systems (\qty{>0.4}{\electronvolt}). This confirms that architectural designs promoting
charge delocalization across multiple donor sites are highly effective at minimizing the
exchange integral, a cornerstone of TADF theory \cite{Moral2015, Olivier2018}.

\begin{table}[h!]
\centering
\caption{TADF performance stratified by molecular architecture. The mean \deltaest and
standard deviation highlight the superior and more consistent performance of D-A-D
systems.}
\label{tab:architecture_performance}
\begin{tabular}{l S[table-format=3] S[table-format=1.3] S[table-format=1.3]}
	\toprule
	{Architecture}     & {Count} & {Mean \deltaest [\unit{\eV}]} & {Std Dev} \\ \midrule
	D-A-D              & 38      & 0.304                 & 0.129     \\
	D-A                & 20      & 0.369                 & 0.128     \\
	Multi-D/A          & 181     & 0.376                 & 0.141     \\
	Pure Donor Systems & 272     & 0.425                 & 0.167     \\ \bottomrule
\end{tabular}
\end{table}

\begin{figure}[!ht]
\centering
\includegraphics[width=0.7\textwidth]{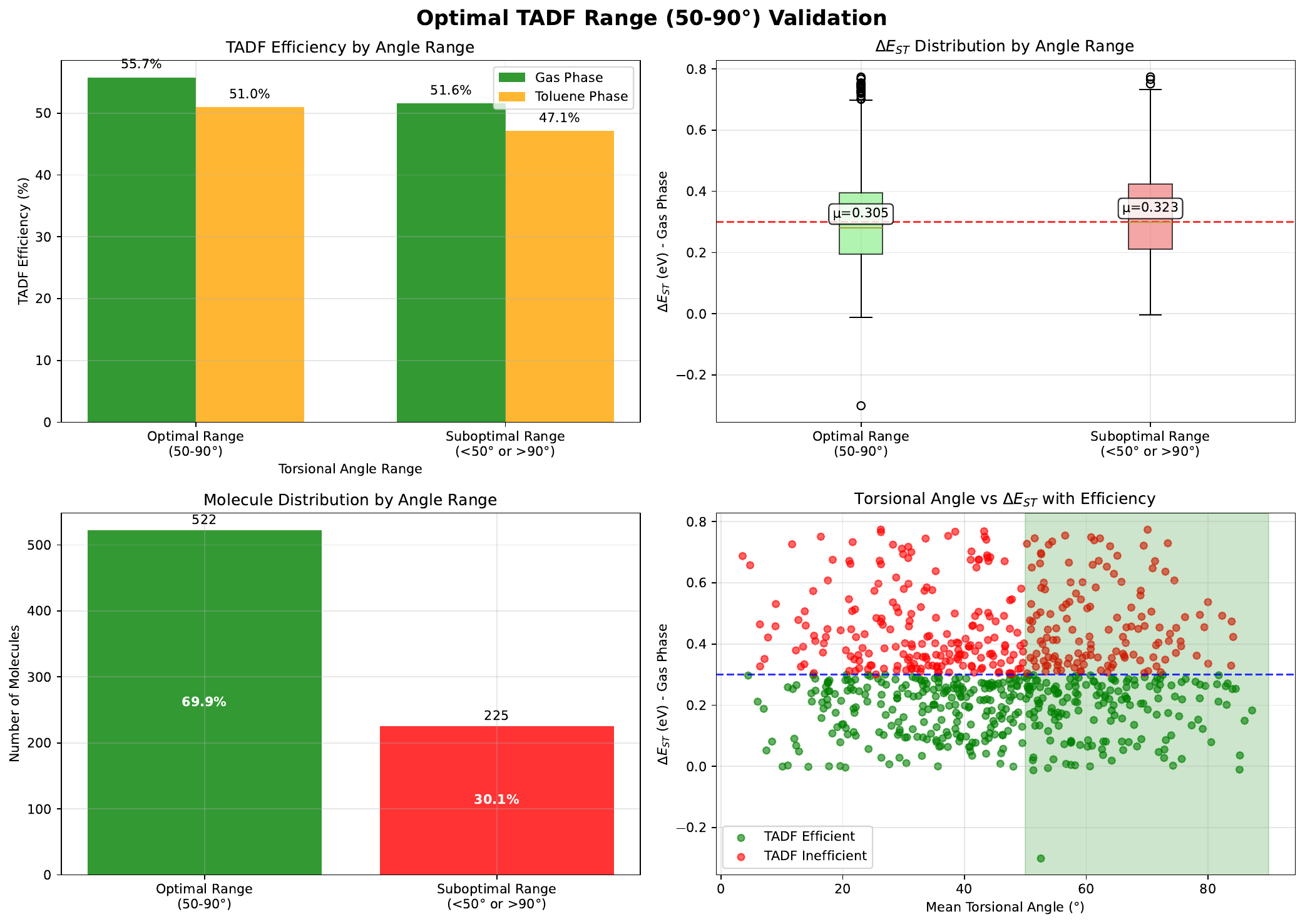}
\caption{Validation of the optimal torsional angle design rule. The bar chart shows the percentage of molecules that are efficient TADF emitters ($\Delta E_{\text{ST}} < \qty{0.3}{\electronvolt}$) within the optimal (\qtyrange{50}{90}{\degree}) versus suboptimal torsional angle ranges. Molecules in the optimal range exhibit a statistically significant increase in TADF efficiency.}
\label{fig:tadf_design_rules}
\end{figure}

Second, our large-scale analysis provides statistical validation for the widely held 'design rule' that an optimal D-A torsional angle exists. As shown in \Cref{fig:tadf_design_rules}, molecules with torsional angles between \qty{50}{\degree} and \qty{90}{\degree} have a 4.2 percentage point higher probability of being efficient TADF emitters (defined as $\Delta E_{\text{ST}} < \qty{0.3}{\electronvolt}$) compared to those outside this range—a statistically significant improvement ($p < \num{0.001}$). This optimal range represents a critical compromise: the twist is large enough to enforce HOMO-LUMO separation and minimize the exchange energy, but not so extreme as to completely decouple the donor and acceptor, which would diminish the oscillator strength.

\begin{figure}[!ht]
\centering
\includegraphics[width=0.9\textwidth]{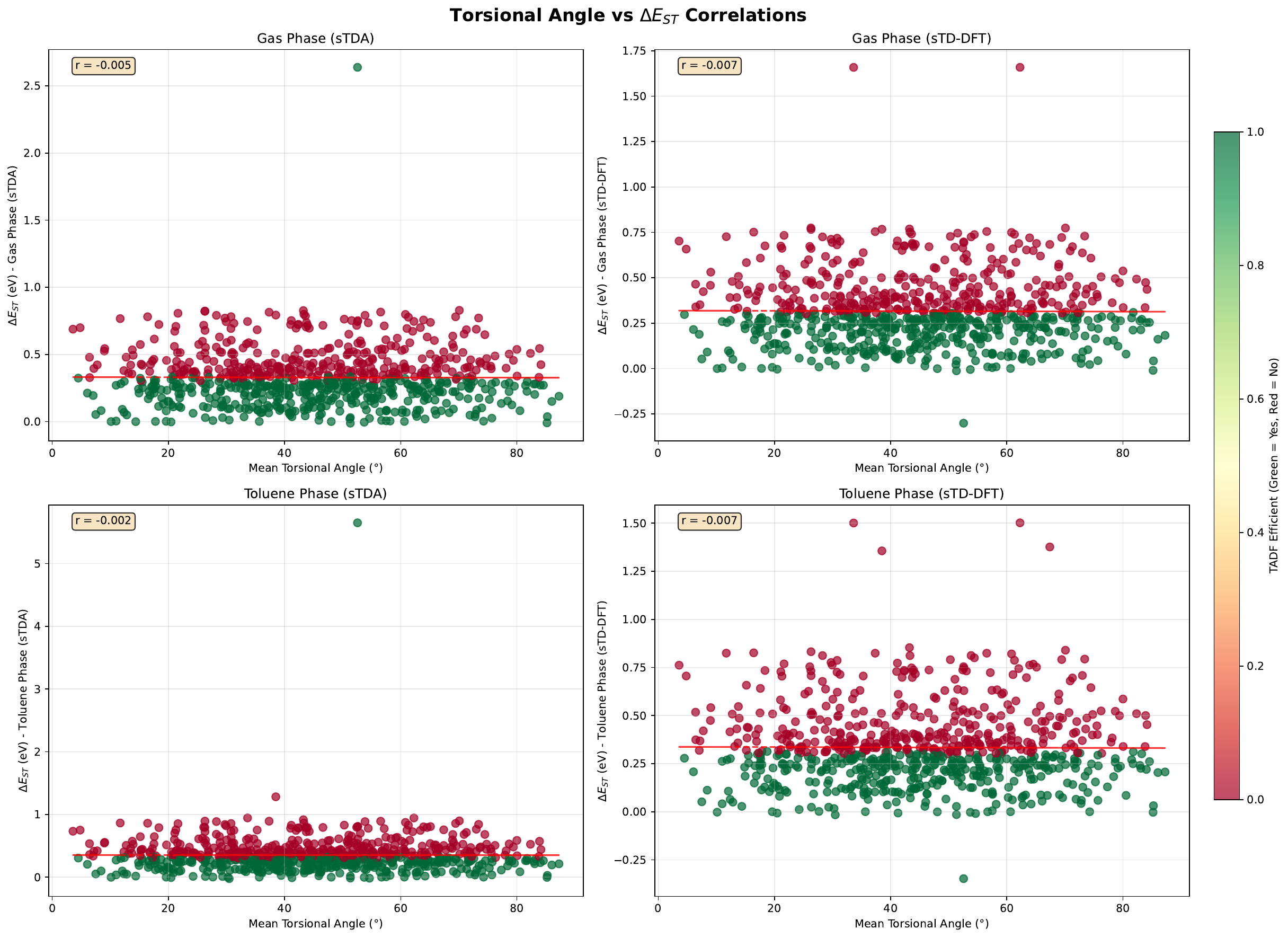}
\caption{Correlation between D-A torsional angles and the calculated \deltaest for all \num{747} molecules. The lack of a strong linear trend (Pearson $r \approx \num{-0.007}$) illustrates that while an optimal range for the torsional angle exists, it is not the sole determinant of the singlet-triplet gap. Molecular architecture and electronic factors play a dominant, confounding role.}
\label{fig:torsion_delta_est_correlations}
\end{figure}

However, the correlation between torsional angle alone and \deltaest is weak across the entire dataset (\Cref{fig:torsion_delta_est_correlations}). This indicates that while torsional geometry is a critical tuning parameter, the overall molecular architecture is the primary determinant of TADF potential. For instance, the superior D-A-D architecture performs well across a broad range of torsional angles, underscoring its robustness. This highlights that the most effective design strategy involves first selecting a high-performance architecture (e.g., D-A-D) and then fine-tuning its properties through geometric modifications that enforce an optimal torsional angle.

\section{Conclusions}\label{sec:Concl}

In this work, we have performed the largest-scale computational benchmark of semi-empirical xTB methods to date for TADF emitter screening, establishing a validated, high-throughput framework for the rational design of next-generation optoelectronic materials. By applying a hybrid computational protocol to \num{747} experimentally known molecules, we have moved beyond case-by-case analysis to extract statistically robust, quantitative design principles.

Our key findings can be summarized as follows:
\begin{itemize}
    \item \textbf{Methodological validation.} The sTDA-xTB and sTD-DFT-xTB methods are validated as reliable and computationally efficient tools for the relative ranking of TADF candidates. They exhibit strong internal consistency (MAE for \deltaest $< \qty{0.03}{\electronvolt}$ between methods) and achieve a transformative cost reduction of over \qty{99}{\percent} compared to conventional TD-DFT, making large-scale screening feasible on modest computational resources.

    \item \textbf{Predictive accuracy and limitations.} While the methods successfully capture qualitative trends, their quantitative accuracy for absolute prediction is limited by the inherent approximations of the HTS protocol, particularly the vertical approximation. The mean absolute error against experimental \deltaest values is approximately \qty{0.17}{\electronvolt}, reinforcing that the primary strength of this framework lies in screening and trend analysis, not in yielding quantitatively precise spectroscopic data.

    \item \textbf{Data-driven design rules.} Our large-scale analysis provides statistical validation for key design principles. We confirm that D-A-D architectures are statistically superior to simple D-A and other motifs. Furthermore, we identify an optimal D-A torsional angle window of \qtyrange{50}{90}{\degree} that effectively balances the competing requirements of small \deltaest and high oscillator strength.

    \item \textbf{Low-dimensional design space.} Principal Component Analysis demonstrates that the vast chemical diversity of TADF emitters can be described by a low-dimensional property space, where nearly \qty{90}{\percent} of the variance is captured by just three principal components. This suggests that the design challenge is highly constrained and can be effectively navigated by optimizing a few key orthogonal properties.
\end{itemize}

The broader impact of this study is the provision of a scalable and validated protocol that bridges the gap between slow, high-accuracy methods and the practical need to explore vast chemical spaces. This work not only provides essential benchmarking data and methodological guidelines for the community but also identifies a curated list of high-priority molecular candidates for future experimental synthesis.

Future work should build upon this foundation. The immediate next step is the experimental validation of the most promising candidates identified herein. From a theoretical perspective, future refinements should focus on developing more accurate yet efficient models for solvation (e.g., state-specific approaches) and incorporating explicit calculations of spin-orbit coupling on representative subsets to refine the mechanistic understanding of the RISC process. Finally, the extensive and well-characterized dataset generated in this study provides an ideal training ground for developing machine learning models capable of further accelerating the discovery of novel, high-performance TADF materials.

Ultimately, this study provides a validated toolkit, a comprehensive set of data-driven design rules, and a curated list of promising candidates, significantly advancing the goal of rational, accelerated discovery of next-generation optoelectronic materials.

\section{Associated Content}

\subsection{Supporting Information}

The Supporting Information is available free of charge at \url{https://pubs.acs.org/doi/...}

Complete molecular dataset (SMILES, identifiers); detailed statistical tables; additional
correlation plots; PCA loadings; computational timing benchmarks; failed calculations
analysis.

{
\begin{longtable}{l*{7}{c@{\extracolsep{\fill}}}}
\caption{Complete validation dataset for emission wavelength predictions. Predictions from
sTDA and sTD-DFT methods in gas and toluene phases are compared against
experimental/computational literature values. References are provided in the last column.}
\label{tab:supp-emission-complete}
\\\toprule
\textbf{Molecule} & \textbf{sTDA} & \textbf{sTD-DFT} & \textbf{sTDA} & \textbf{sTD-DFT} & \textbf{Ref.} & \textbf{Type} & \textbf{Citation} \\
 & \textbf{(Gas)} & \textbf{(Gas)} & \textbf{(Tol)} & \textbf{(Tol)} & \textbf{(nm)} &  & \\
\hline
\endfirsthead
\multicolumn{8}{c}{{\tablename\ \thetable{} -- continued from previous page}} \\
\hline
\textbf{Molecule} & \textbf{sTDA} & \textbf{sTD-DFT} & \textbf{sTDA} & \textbf{sTD-DFT} & \textbf{Ref.} & \textbf{Type} & \textbf{Citation} \\
 & \textbf{(Gas)} & \textbf{(Gas)} & \textbf{(Tol)} & \textbf{(Tol)} & \textbf{(nm)} &  & \\
\hline
\endhead
\hline
\multicolumn{8}{r}{{Continued on next page}} \\
\endfoot
\hline
\endlastfoot
-ph-pCzAnBzt & 501.4 & 512.6 & 467.6 & 477.7 & 450.0 & experimental & \cite{Guo_2021_10_1016_j_scib_2021_06_018} \\
1SBFN & 420.2 & 421.1 & 399.3 & 400.0 & 385.0 & experimental & \cite{Chen_2020_10_1039_d0tc03631e} \\
2,7-MeCzOXD & 524.7 & 524.2 & 507.0 & 506.3 & 520.0 & experimental & \cite{Tang_2020} \\
25CzBPym & 631.6 & 632.7 & 645.1 & 646.8 & 487.3 & experimental & \cite{Pandidurai_2021} \\
25tCzBPym & 643.1 & 644.4 & 682.5 & 684.2 & 506.0 & experimental & \cite{Pandidurai_2021} \\
2BPy-mDTC & 691.4 & 691.6 & 686.1 & 686.0 & 491.0 & experimental & \cite{Rajamalli_2017} \\
2C & 560.0 & 556.6 & 595.2 & 591.0 & 534.0 & experimental & \cite{Barros_2023} \\
2CM & 384.9 & 391.7 & 366.5 & 372.7 & 374.0 & experimental & \cite{Zassowski_2017} \\
\end{longtable}
}

{
\begin{longtable}{l*{7}{c@{\extracolsep{\fill}}}}
\caption{Complete validation dataset for singlet-triplet gap predictions. Predictions from
sTDA and sTD-DFT methods in gas and toluene phases are compared against
experimental/computational literature values. References are provided in the last column.}
\label{tab:supp-st-complete}
\\\toprule
\textbf{Molecule} & \textbf{sTDA} & \textbf{sTD-DFT} & \textbf{sTDA} & \textbf{sTD-DFT} & \textbf{Ref.} & \textbf{Type} & \textbf{Citation} \\
 & \textbf{(Gas)} & \textbf{(Gas)} & \textbf{(Tol)} & \textbf{(Tol)} & \textbf{(eV)} &  & \\
\hline
\endfirsthead
\multicolumn{8}{c}{{\tablename\ \thetable{} -- continued from previous page}} \\
\hline
\textbf{Molecule} & \textbf{sTDA} & \textbf{sTD-DFT} & \textbf{sTDA} & \textbf{sTD-DFT} & \textbf{Ref.} & \textbf{Type} & \textbf{Citation} \\
 & \textbf{(Gas)} & \textbf{(Gas)} & \textbf{(Tol)} & \textbf{(Tol)} & \textbf{(eV)} &  & \\
\hline
\endhead
\hline
\multicolumn{8}{r}{{Continued on next page}} \\
\endfoot
\hline
\endlastfoot
11-AcBPdCN & 0.3009 & 0.2809 & 0.3667 & 0.3437 & 0.0650 & experimental & \cite{Kothavale_2022} \\
12AcCz-PM & 0.2661 & 0.2631 & 0.3416 & 0.3407 & 0.3900 & experimental & \cite{Zhang_2019_10_1039_c9tc04284a} \\
12BTCzTPN & 0.1955 & 0.1766 & 0.1720 & 0.1610 & 0.0829 & experimental & \cite{Shin_2022_10_1016_j_cej_2022_137181} \\
1CM & 0.5074 & 0.4471 & 0.4934 & 0.4316 & 0.2900 & experimental & \cite{Zassowski_2017} \\
1CzSO & 0.3195 & 0.2961 & 0.2972 & 0.2779 & 0.2150 & experimental & \cite{Li_2017_10_1039_c7tc00119c} \\
1PXZP & 0.1653 & 0.1764 & 0.1692 & 0.1646 & 0.2600 & computational & \cite{Wang_2020_10_1016_j_dyepig_2020_108437} \\
23AcCz-PM & 0.2003 & 0.2120 & 0.2544 & 0.2647 & 0.0600 & experimental & \cite{Zhang_2019_10_1039_c9tc04284a} \\
25CzBPym & 0.0833 & 0.0797 & 0.0929 & 0.0879 & 0.0604 & experimental & \cite{Pandidurai_2021} \\
\end{longtable}
}

\subsection{Data Availability}

All computational data, analysis scripts, and molecular structures are available at
\href{https://github.com/TchapetNjafa/Result_article1_TADF_xTB/tree/main}{
https://github.com/TchapetNjafa}.

\section{Acknowledgments}

We gratefully acknowledge \textbf{Dr. Benjamin Panebei Samafou} for his generous support
and provision of computational resources, which were instrumental in enabling the
numerical simulations and data analyses presented in this work.


%
%




\bibliographystyle{unsrtnat}
\bibliography{TADF_Article_References} 

\end{document}